\documentclass[useAMS]{mn2e}
\usepackage{graphicx,amssymb,amsmath}
\usepackage{hyperref,xcolor}
\usepackage{subfigure}
\usepackage{pdflscape}
\usepackage{mathrsfs}
\newcommand{\scs}{\scriptsize}

\title[Argus, Carina-Near, Hercules-Lyra, Orion and Subgroup B4] {Local associations and the barium puzzle }
\author[A. B. S. Reddy and D. L. Lambert] { Arumalla B. S. Reddy$^{1}$\thanks{E-mail: [bala; dll]@astro.as.utexas.edu} and David L. Lambert$^1$ \\
  $^1$W.J. McDonald Observatory and Department of Astronomy, The University of Texas at Austin, Austin, TX 78712-1205, USA }
\begin{document}

\pagerange{\pageref{firstpage}--\pageref{lastpage}} \pubyear{2015}

\maketitle

\label{firstpage}

\begin{abstract}

We have observed high-dispersion echelle spectra of main-sequence stars in five nearby young associations -- Argus, Carina-Near, Hercules-Lyra, Orion and Subgroup B4 -- and derived abundances for elements ranging from Na to Eu. These are the first chemical abundance measurements for two of the five associations, while the remaining three associations are analysed more extensively in our study. Our results support the presence of chemical homogeneity among association members with a typical star-to-star abundance scatter of about 0.06 dex or less over many elements. The five associations show log\,$\epsilon$(Li) consistent with their age and share a solar chemical composition for all elements with the exception of Ba. We find that all the heavy elements (Y, Zr, La, Ce, Nd, Sm and Eu) exhibit solar ratios, i.e., [X/Fe] $\simeq$ 0, while Ba is overabundant by about 0.2$-$0.3 dex. 
The origin of the overabundance of Ba is a puzzle. Within the formulation of the $s$-process, it is difficult to create a higher Ba abundance without a similar increase in the $s$-process contributions to other heavy elements (La-Sm). Given that Ba is represented by strong lines of Ba\,{\sc ii} and  La-Sm are represented by rather weak ionized lines, the suggestion, as previously made by other studies, is that the Ba abundance may be systematically overestimated by standard methods of abundance analysis perhaps because the upper reaches of the stellar atmospheres are poorly represented by standard model atmospheres. A novel attempt to analyse the Ba\,{\sc i} line at 5535 \AA\ gives a solar Ba abundance for stars with effective temperatures hotter than about 5800 K but increasingly subsolar Ba abundances for cooler stars with apparent Ba deficiencies of 0.5 dex at 5100 K. This trend with temperature may signal a serious non-LTE effect on the Ba\,{\sc i} line.

\end{abstract}


\section{Introduction} 

Over the last few years, several comprehensive high-resolution studies of chemical abundances in the solar vicinity have been reported for field dwarfs (Edvardsson et al. 1993; Reddy et al. 2003, 2006; Luck \& Heiter 2005, 2006; Bensby et al. 2005, 2014), field giants (Mishenina et al. 2006; Luck \& Heiter 2007; Takeda et al. 2008), and for open clusters (OCs; Friel et al. 2002; Bragaglia et al. 2008; Yong et al 2012; Reddy et al. 2013, 2015), with the main goal of improving our knowledge of the chemical evolution of the Galactic disc.

The results suggest that both OCs and field stars have very similar chemical compositions, i.e., the same [X/Fe] for the element X, at a given metallicity and age (Bragaglia et al. 2008; Yong et al. 2012; Reddy et al. 2015). This is a pleasing result because similar compositions are expected 
as dissolving OCs are believed to be the principal source of field stars (Lada \& Lada 2003). However, the pleasing result has been disturbed by reports of abundance differences between OCs and field stars for heavy elements. In particular, D'Orazi et al. (2009) from analysis of the Ba\,{\sc ii} 5853 \AA\ and 6496 \AA\ lines in dwarf stars from 20 OCs with ages from 0.04 Gyr to 8.4 Gyr found [Ba/Fe] $\simeq +0.6$ in the youngest clusters with [Ba/Fe] decreasing to $\simeq 0.0$  in OCs of ages about two Gyr and older. (Giants in the same clusters gave [Ba/Fe] values about $+0.2$ larger.) These Ba abundances for stars in OCs are seemingly at odds with results obtained from field dwarfs and giants.  A couple of points deserve a mention. 

\begin{table*}
\centering
\caption{Targeted Stellar associations and their properties from the literature.}
\label{tab1}
\begin{tabular}{lccrccccc}   \hline
\multicolumn{1}{l}{Association}&  \multicolumn{1}{c}{Distance}& \multicolumn{1}{c}{Age}& \multicolumn{1}{r}{Star name} & 
\multicolumn{1}{l}{$\alpha(2000.0)$}& \multicolumn{1}{c}{$\delta(2000.0)$}& \multicolumn{1}{c}{distance of} &
\multicolumn{1}{c}{SpT}& \multicolumn{1}{c}{Ref$^{a}$} \\
\multicolumn{1}{c}{}& \multicolumn{1}{c}{ (pc) }& \multicolumn{1}{c}{(Myr)}& \multicolumn{1}{c}{ } &
\multicolumn{1}{l}{(hh mm s)}& \multicolumn{1}{c}{($\degr$ $\arcsec$ $\arcmin$)} & \multicolumn{1}{c}{star (in pc)}& \multicolumn{1}{c}{ }&
\multicolumn{1}{l}{ } \\  \hline

Argus         & 65 & 30  & BD-20 2977 & 09 39 51.42 & $-$21 34 17.45 & 88.5 & G9V & [1] \\
              &    &     &   HD 61005 & 07 35 47.46 & $-$32 12 14.04 & 35.3 & G8V & [1] \\
Carina-Near   & 30 & 200 &  HD 103742 & 11 56 42.31 & $-$32 16 05.41 & 34.9 & G3V & [2] \\
              &    &     &  HD 103743 & 11 56 43.76 & $-$32 16 02.70 & 29.2 & G4V & [2] \\
Hercules-Lyra & 20 & 200 &     HD 166 & 00 06 36.78 & $+$29 01 17.40 & 13.7 & K0V & [3] \\
              &    &     &   HD 10008 & 01 37 35.47 & $-$06 45 37.53 & 23.6 & G5V & [3] \\
              &    &     &   HD 17925 & 02 52 32.13 & $-$12 46 10.97 & 10.4 & K1V & [3] \\
Orion (sub-group Ic) & 50 &  3  &Parenago 2339 & 05 36 06.05 &$-$06 19 38.94& $\sim$ 50 & F9IV & [4] \\
              &    &     &Parenago 2374 & 05 36 15.23 &$-$06 23 55.65& $\sim$ 50 & F9IV & [4] \\
Subgroup B4   & 32 & 80  &  HD 113449 & 13 03 49.65 & $-$05 09 42.52 & 22.1 & K1V & [3] \\
              &    &     &  HD 152555 & 16 54 08.14 & $-$04 20 24.66 & 47.6 & G0V & [3] \\   
\hline
\end{tabular}
\flushleft
$^{a}$ Reference for the target selection: [1] De Silva et al. (2013); [2] Zuckerman et al. (2006); [3] L\'{o}pez-Santiago et al. (2006); [4] Cunha et al. (1995)  
\end{table*}

Studies of field stars have not shown a tendency for Ba abundances to increase sharply with decreasing age. An early study (Edvardsson et al. 1993) suggested that at [Fe/H] $\sim 0.0$ field FG dwarfs with ages less than about 4 Gyr had [Ba/Fe] $\simeq +0.05$  but older stars had a lower [Ba/Fe]  with [Ba/Fe] $\simeq -0.1$ for ages between 4 and 8 Gyr. In a recent study, da Silva et al. (2015) found [Ba/Fe] $\simeq$ 0.0 for 120 field dwarfs from the thin disc with a slight rise to no more that $+0.1$ at 0.5 Gyr for a sample spanning the [Fe/H] range $+0.2$ to $-0.5$.  Although in terms of age distribution the OCs and field show little overlap at the youngest ages, it seems clear that stars with high Ba overabundances such as  [Ba/Fe] $\simeq +0.6$, as reported for young OCs, are not represented among field stars.  A suggestion that the young OCs have yet to dissolve and feed the field star population is not appealing as it implies an abrupt recent and large increase in evolution of the $s$-process yields for the Galactic disc. (The few field dwarf and giants stars with [Ba/Fe] $\gg 0.0$ are known as Barium stars and are binaries in which an AGB companion, now present as a white dwarf, has transferred $s$-process rich material to create the Barium star.)

Another indicator of apparently different Ba abundances in OCs and field stars was provided by Mishenina et al.'s (2015) measurements and compilation of Ba abundances for OCs with [Fe/H] from  about  $-0.5$ to $+0.2$. At ages of less than about 1 Gyr, the OCs show a [Ba/Fe] range from about 0.0 to $+0.7$ from a mix of dwarfs and giants and in the published figure (their Figure 4) the [Ba/Fe] for field stars extrapolated to young ages is approximately 0.0 to $+0.1$ (see previous paragraph). A majority of the results for the few older OCs are in the range [Ba/Fe] $+0.3$ to about $+0.5$ but all field stars have [Ba/Fe] $\simeq 0.0$. These differences for Ba between OCs and field stars are smaller to non-existent for Y and La (see Mishenina et al. 2015, their Figures 3 and 4) and also for Zr, Nd, Sm and Eu (Reddy et al. 2015). For the older OCs, one expects a close correspondence in composition between surviving clusters and field stars which in the main are expected to have come from similarly-aged OCs. The OC-field star difference for Ba  with the lack of a difference for other heavy elements in the interval Y-Eu suggests the resolution of the Barium puzzle lies in an incomplete understanding of the formation of strong Ba\,{\sc ii} lines in spectra of F-G-K dwarfs and giants rather than in unanticipated aspects of the chemical evolution of OCs, especially younger clusters.

Another dimension to the Barium puzzle was added by D'Orazi et al. (2012) who analysed the heavy elements Y, Zr, Ba, La, and Ce in FG dwarfs in three local associations with ages of less than 500 Myr. These elements except for Ba had solar [X/Fe] values but Ba was overabundant with [Ba/Fe] $\simeq +0.2$, a value appreciably less than the $+0.6$ reported in D'Orazi et al. (2009) for the youngest OCs. Noting that the ratio of the Ba abundance to that of other heavy elements is not easily replicated by operation of the $s$-process, D'Orazi et al. (2012) speculated that the strong Ba\,{\sc ii} lines - the source of the Ba abundance - are prone to atmospheric effects not accounted for in the standard LTE analysis; possible effects include `the presence of hot stellar chromospheres'. This study was followed by De Silva et al. (2013) who derived the Ba abundance for six stars in the Argus association of age 30 Myr finding a mean [Ba/Fe] of $+0.53\pm0.08$ and eight stars in the young OC IC 2391 gave [Ba/Fe]$= +0.62\pm0.07$. The chemical content of Argus association was measured previously for one star HD 61005 by Desidera et al. (2011) who derived a mean [Ba/Fe] of $0.63\pm0.06$. Unfortunately, no other heavy elements were considered by Desidera et al. (2011) and De Silva et al. (2013) who returned the Ba abundance in associations to the high level reported earlier by D'Orazi et al. (2009) for the youngest OCs.

To our knowledge, only the four associations just mentioned have been analysed for the abundance of at least one heavy element (i.e., Ba). In this paper, we expand the information available on heavy elements in young associations. In particular, we consider the suite of elements Y, Zr, Ba, La, Ce, Nd, Sm and Eu with several elements and notably Eu considered for the first time in definition of the composition of associations. Additionally, our analysis is comprehensive in that we cover elements from Li to Zn in addition to the heavy elements. Two of the sample of five associations -- Hercules-Lyra and Subgroup B4 -- are analysed for the first time. For the other three associations, we either provide a more extensive analysis of stars previously analysed or the first analysis of a member star. The associations selected for abundance analysis included: Argus (De Silva et al. 2013), Carina-Near (Zuckerman et al. 2006), Hercules-Lyra and Subgroup B4 (L\'{o}pez-Santiago et al. 2006), and Orion (Cunha et al. 1995). Our overall goal is to improve our understanding of chemical evolution in the solar vicinity. We compare our newly observed associations with those in the literature, and with stars in the field and the young OCs in the Galactic disc. 

The layout of the paper is as follows: In Section 2 we describe the target selection and observations, and Section 3 is devoted to the data reduction and radial velocity measurements. In Section 4 we discuss the abundance analysis and in Section 5 we present our results with a short discussion on each of the association analysed here with a summary subsection. Section 6 is discusses the barium abundance in relation to other $s$-process elements and to elements of other species represented by lines of very similar strengths to the barium lines. We discuss in Section 7 the relative abundances of associations in comparison with stars in OCs and the field. Finally, in Section 8 we provide our conclusions concerning the central questions about the Ba and other heavy element abundances in local stellar associations. 

\begin{figure*}
\begin{center}
\includegraphics[trim=0.10cm 0.0cm 0.95cm 5.4cm, clip=true,height=0.36\textheight,width=0.68\textheight]{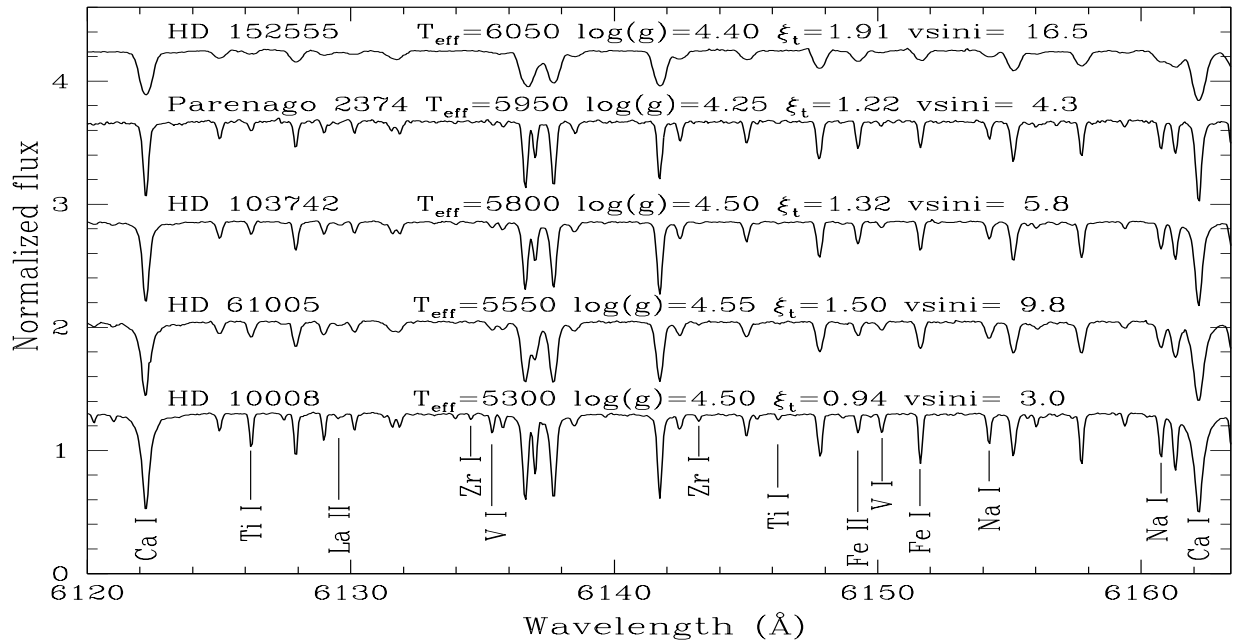} \vspace{-0.2cm}
\caption{Normalized spectra of dwarf members of the five young associations as described in Table \ref{tab2} are presented in descending order of temperature (top to bottom) in the 6120$-$6165 \AA\ region. All the spectra are on the same scale but have been displaced vertically from each other for clarity.}
\label{spectra} 
\end{center}
\end{figure*}
  
\section{Target selection and Observations}

The sample of low mass stars was selected from the list of high probability members of nearby young stellar associations -- Argus, Carina-Near, Hercules-Lyra, Orion and Subgroup B4. The chemical composition of Argus association was previously analysed for the elements Si, Ni and Fe by Viana Almeida et al. (2009, hereafter VA09), Na, Si, Ca, Ti, Ni, Fe and Ba by Desidera et al. (2011), and for the elements Na, Al, $\alpha$-elements (Mg, Si, Ca, and Ti), iron-peak elements (Cr, Ni, and Fe), and the $s$-process element (Ba) by De Silva et al. (2013). The chemical content of Carina-Near was measured previously by Biazzo et al. (2012) for elements from Li to Zn and by D'Orazi et al. (2012) for the heavy elements (Y, Zr, Ba, La, and Ce). Our analysis of Carina-Near which  included two stars (HD 103742 and HD 103743) in common with their work offers an opportunity to assess the systematic differences in elemental abundances among the analyses. The Orion association has been analysed previously for the elements Li and Fe by Cunha et al. (1995), and for O  and Fe by Cunha et al. (1998) for a sample of F and G dwarfs. The chemical content of Hercules-Lyra and Subgroup B4 group members has been explored for the first time in this study. 

In this paper, we have performed a homogeneous and a comprehensive abundance analysis using high resolution spectroscopy. We employed strict criteria in the selection of suitable stars for abundance analysis. We selected only dwarf stars with spectral types from late-F through G to early K with projected rotational velocities $v$~sin$i \leq$ 20 km s$^{-1}$ and having no sign of spectroscopic binarity. Stars with later spectral types were excluded as their atmospheres produce significant molecular bands at effective temperatures lower than $\sim$ 4000 K thereby yielding unreliable chemical compositions through equivalent width measurements. FGK stars with high $v$~sin$i$ were neglected primarily because the weak photospheric lines, especially due to the $s$- and $r$-process elements are smeared out and disappear completely into the continuum and secondarily stronger lines are often significantly blended by rotational broadening. 

\begin{table*}
%  \centering
% {\fontsize{6}{8}\selectfont 
 \caption{The journal of the observations for each of the association members analysed in this paper.} 
 \label{tab2}
\begin{tabular}{lcccccccc}   \hline
\multicolumn{1}{l}{Star}& \multicolumn{1}{c}{V}& \multicolumn{1}{c}{(B-V)} & \multicolumn{1}{c}{(V-K$_{\rm s}$)} & \multicolumn{1}{l}{(J-K$_{\rm s}$)} & \multicolumn{1}{l}{Date of} & \multicolumn{1}{c}{Exp. time} & \multicolumn{1}{l}{S/N at} & \multicolumn{1}{c}{$RV_{\rm helio}$}  \\
\multicolumn{1}{l}{ }& \multicolumn{1}{c}{(mag.)}& \multicolumn{1}{c}{ } & \multicolumn{1}{c}{ } & \multicolumn{1}{l}{ } & \multicolumn{1}{l}{observation} & \multicolumn{1}{c}{(sec.)} & \multicolumn{1}{l}{6000 \AA } & \multicolumn{1}{c}{(km s$^{-1}$)}  \\
\hline

BD-20 2977   & 10.21 &$+$0.72 & 1.83 &  0.45 & 2015 Feb 11 & 2$\times$1800 & 140 &$+$19.6$\pm$0.4 \\
 HD 61005    & 08.22 &$+$0.74 & 1.76 &  0.45 & 2015 Feb 11 & 1$\times$1500 & 220 &$+$22.9$\pm$0.3 \\
HD 103742    & 07.64 &$+$0.64 & 1.51 &  0.36 & 2015 Feb 11 & 2$\times$1200 & 370 &$+$07.3$\pm$0.2 \\
HD 103743    & 07.81 &$+$0.67 & 1.57 &  0.36 & 2015 Feb 11 & 2$\times$1800 & 270 &$+$08.2$\pm$0.2 \\
   HD 166    & 06.13 &$+$0.75 & 1.82 &  0.42 & 2015 Feb 11 & 1$\times$180  & 200 &$-$06.4$\pm$0.1 \\
 HD 10008    & 07.66 &$+$0.80 & 1.91 &  0.47 & 2015 Feb 11 & 1$\times$1200 & 230 &$+$11.8$\pm$0.1 \\
 HD 17925    & 06.05 &$+$0.86 & 1.98 &  0.55 & 2015 Feb 11 & 1$\times$300  & 220 &$+$18.2$\pm$0.1 \\
Parenago 2339& 09.95 &$+$0.51 & 1.32 &  0.26 & 2015 Feb 10 & 2$\times$1800 & 150 &$+$18.6$\pm$0.7 \\
Parenago 2374& 11.09 &$+$0.53 & 1.67 &  0.34 & 2015 Feb 11 & 2$\times$1800 & 130 &$+$18.6$\pm$0.2 \\
 HD 113449   & 07.68 &$+$0.88 & 2.17 &  0.54 & 2015 Feb 11 & 2$\times$1200 & 200 &$-$09.1$\pm$0.1 \\
 HD 152555   & 07.82 &$+$0.60 & 1.46 &  0.34 & 2015 Feb 11 & 2$\times$1800 & 150 &$-$15.1$\pm$0.4 \\
    
\hline
\end{tabular} 
% }
\end{table*}

We have made use of the {\small SIMBAD}\footnote{\url{http://simbad.u-strasbg.fr/simbad/}} astronomical database for the available astrometric and photometric measurements of the association members and the more recent values were adopted. The five associations are young and in the immediate neighbourhood of the Sun: the heliocentric distances run from 20 to 65 pc and the ages range from 3 to 200 Myr (see Table \ref{tab1}).

High-resolution and high signal-to-noise (S/N) spectra of the program stars were obtained during the nights of 2015 February 10-11 with the Robert G. Tull coud\'{e} cross-dispersed echelle spectrograph (Tull et al. 1995) of the 2.7-m Harlan J. Smith reflector at the McDonald observatory. We employed a Tektronix 2048$\times$2048 24 $\mu$m pixel, backside illuminated and anti-reflection coated CCD as a detector and a R2 echelle grating with 52.67 grooves mm$^{-1}$ with exposures centred at 5060 \AA.  

Each night's observing routine included five zero second exposures (bias frames), 15 quartz lamp exposures (flat frames), and finally two Th-Ar hollow cathode spectra which provide the reference wavelength scale. A total of eleven stars spread across the five associations were observed. For each target, we obtained two exposures each lasting for 5-30 min to minimize the influence of cosmic rays and to acquire a good S/N ratio. All the spectra, with the exception of Parenago 2339, correspond to a resolving power of $R$ $\gtrsim$ 55,000 ($<$ 6 km s$^{-1}$) as measured by the FWHM of Th {\scs I} lines in comparison spectra, while the latter spectra were taken at a lower resolution of about 30,000. The spectral coverage in a single exposure from 3600 \AA\ to 9800 \AA\ across various orders was complete but for the inter-order gaps beyond $\sim$ 5600 \AA\ where the echelle orders were incompletely captured on the CCD. 

\section{Data reduction and radial velocities}

The spectroscopic data reductions were performed in multiple steps using various routines available within the \textit{imred} and \textit{echelle} packages of the standard spectral reduction software {\scs IRAF}\footnote{IRAF is a general purpose software system for the reduction and analysis of astronomical data distributed by NOAO, which is operated by the Association of Universities for Research in Astronomy, Inc. under cooperative agreement with the National Science Foundation.}. Briefly, the two-dimensional target star frames were de-trended by subtracting bias level, correcting for scattered light and then divided by the normalized flat field. The individual echelle orders were traced, extracted to one-dimensional spectra with y-axis as flux and x-axis in pixels and then wavelength calibrated using Th-Ar lamp spectra as a reference. 

Multiple spectra were combined to provide a single, high S/N spectra for each star. The combined spectra have S/N ratios of about 130-370 as measured around 6000 \AA\ region, while at wavelengths shorter than about 4000 \AA\ the S/N ratio decreases with wavelength and reaches a value of about 10 around 3600 \AA\ region. The noisy ends of each echelle order were trimmed to reduce the edge effects before continuum fitting. The spectrum was normalized interactively to unity with the available cursor commands in \textit{splot} task of {\scs IRAF} by marking continuum regions, those regions which are uneffected by the presence of absorption lines, on each aperture which were then fitted by a slowly varying function such as cubic spline of appropriate order for better normalization. We referred our spectra to the high-resolution spectrum of Arcturus and the Sun (Hinkle et al. 2000) to identify the continuum regions. Spectra of a representative region are shown in Figure \ref{spectra} for one star from each of the five associations.

The radial velocity (RV) of each star was measured from its normalized spectrum using a set of 20 lines with well defined line cores. The absence of systematic RV shift for the spectral lines selected from the blue and red wavelength regions strengthens the accuracy of our wavelength solution. The observed RVs were corrected for solar motion using the {\it rvcorrect} routine in {\scs IRAF}. Our mean RV measurements for all stars are in fair agreement with the previous RV measurements in the references listed in the last column of Table \ref{tab1}.

The list of program stars and the journal of observations are given in Table \ref{tab2} together with the available optical and 2MASS\footnote{\url{http://irsa.ipac.caltech.edu/applications/Gator}} photometry (Cutri et al. 2003)\footnote{Originally published by the University of Massachusetts and Infrared Processing and Analysis Center (IPAC)/ California Institute of Technology.}, date of observation, exposure time, S/N of the spectra around 6000 \AA\ and computed heliocentric RVs.  

\begin{table*}
{\fontsize{7}{8}\selectfont
\caption{The linelist for all members of associations analysed in this paper and the numbers in columns 5-15 represent the line's EW(m\AA) measured  for each star.}
\label{EWmeasurement}
\begin{tabular}{lcllrrrrrrrrrrr} \hline
\multicolumn{4}{l}{ } & \multicolumn{2}{c}{Argus} \vline& \multicolumn{2}{c}{Carina-Near} \vline& \multicolumn{3}{c}{Hercules-Lyra} \vline& \multicolumn{2}{c}{Orion} \vline& \multicolumn{2}{c}{Subgroup B4}  \\ \cline{5-15}
\multicolumn{1}{l}{$\lambda$(\AA)} & \multicolumn{1}{l}{Species$^{a}$ } & \multicolumn{1}{c}{LEP$^{b}$} & \multicolumn{1}{c}{$\log~{gf}$} & \multicolumn{1}{c}{[1]} & \multicolumn{1}{c}{[2]} \vline& \multicolumn{1}{c}{[3]} & \multicolumn{1}{c}{[4]} \vline&\multicolumn{1}{c}{[5]} & \multicolumn{1}{c}{[6]} & \multicolumn{1}{c}{[7]} \vline& \multicolumn{1}{c}{[8]} & \multicolumn{1}{c}{[9]} \vline& \multicolumn{1}{c}{[10]} & \multicolumn{1}{c}{[11]}  \\ \hline
 
   6707.81 &  3.0 & 0.00 & 0.170 &  242.0 & 166.5 & 114.9 & 111.5 & 77.5 & 99.8 & 205.8 & 109.3 & 45.0 & 148.2  & 136.2 \\
   7771.97 &  8.0 & 9.15 & 0.364 &   58.7 &  52.3 &  73.1 &  71.2 & 55.8 & 41.4 &  35.0 &  95.0 & 93.0 &  31.0  &  88.8 \\
   7774.13 &  8.0 & 9.15 & 0.223 &$\ldots$&  46.8 &  68.4 &  60.7 & 46.8 & 33.0 &  28.0 &  82.4 & 80.9 &$\ldots$&  76.7 \\
   7775.41 &  8.0 & 9.15 & 0.002 &   39.5 &  36.9 &  52.4 &  50.8 & 34.4 & 24.8 &  21.8 &  72.4 & 64.3 &  21.7  &  62.3 \\
 
\hline
\end{tabular} }
\flushleft  
[1]=BD-20 2977, [2]=HD 61005; [3]=HD 103742, [4]=HD 103743; [5]=HD 166, [6]=HD 10008, [7]=HD 17925; [8]=Parenago 2339, [9]=Parenago 2374; [10]= HD 113449, [11]=HD 152555. \\
$^{a}$ The integer part of the 'Species' indicates the atomic number, and the decimal component indicates the ionization state \\ 
\hspace{0.2cm} (0 = neutral, 1 = singly ionized). \\
$^{b}$ All the lines are arranged in the order of their increasing Lower Excitation Potential (LEP). \\ 
Note. Only a portion of this table is shown here for guidance regarding its form and content. A machine-readable version of the full table is available as Supporting Information with the online version of this Paper. 
\end{table*}

\section{Abundance analysis} 

\subsection{\small Line selection and Equivalent widths} \label{line_selection}

We followed the basic local thermodynamical equilibrium (LTE) abundance analysis procedure as described in Reddy et al. (2012), so we refer the reader to that paper for more information on the adopted initial line list and the source of references to atomic data. We augmented the 2012 line list by adding lines due to Li {\scs I}, O {\scs I} triplet and a few additional Fe {\scs I} and Fe {\scs II} lines. The hfs data for the Li {\scs I} has been taken from Reddy et al. (2002). The atomic data for the oxygen triplet has been taken from {\scs \bf NIST}\footnote{\url{http://physics.nist.gov/PhysRefData/ASD/lines$\_$form.html}} database. We constructed our line list with clean, unblended, isolated and symmetric weak and moderately strong lines avoiding the wavelength regions affected by telluric contamination and heavy line crowding. The line equivalent widths (EWs) were measured manually using the cursor commands in \textit{splot} package of {\scs IRAF} by fitting often a Gaussian profile and for a few lines a direct integration was performed for the best measure of EW.     

Our final line list has about 300 absorption lines covering Li and O, as well as 22 elements from Na-Eu in the spectral range $\sim$ 4400 $-$ 8850 \AA. Our selection provides on average across the sample of 11 stars a list of 120 Fe {\scs I} lines with lower excitation potentials (LEPs) ranging from 0.1 to 5.0 eV and EWs of up to 180 m\AA\ and 13 Fe {\scs II} lines with LEPs of 2.8 to 3.9 eV and EWs from $\simeq$ 12 to 90 m\AA. 

As the principal goal is to have the best possible abundance determination for elements represented by few lines we have identified a set of self-consistent absorption lines having high-quality relative $gf$-values. Abundances for the heavy elements except Ba an Eu are based on a combination of very weak (EW $\lesssim$ 10 m\AA) and weak lines (EW $\simeq$ 10$-$60 m\AA). At the resolution of our spectra all these features appear clean, unblended, isolated and the continuum fitting has been done securely around those lines, their EWs can be measured with fair certainty. We derived the Ba abundance using the strong but isolated Ba {\scs II} feature at 5853 \AA, which is quite insensitive to non-LTE effects. The smallest non-LTE correction, of about $\sim -$0.05 dex, found for the 5853 \AA\ line makes it a reliable indicator of the barium abundances in dwarf stars (Korotin et al. 2015). Abundances for the rest of the elements in the list are based either on the EW measurements of weak and moderately strong lines or on spectrum synthesis. We provide in Table \ref{EWmeasurement} the atomic data adopted for each spectral line and the EW measurements for all stars analysed.

In order to minimize the impact of systematic errors in the final stellar abundances, we carried out a strictly differential abundance analysis with respect to the Sun. Solar abundances were derived using solar EWs, measured off the solar integrated disc spectrum (Kurucz et al. 1984), and the ATLAS9 model atmosphere for T$_{\rm eff},_{\odot}$ = 5777 K, log~{g}$_{\odot}$ = 4.44 cm s$^{-2}$ with the overshooting option switched off. We found a microturbulence velocity of $\xi_{t}$ = 0.93 km s$^{-1}$ using both the Fe {\scs I} and Fe {\scs II} lines. Our measured solar abundances reported in Table 4 of Reddy et al. (2012) agree well with the published values of Asplund et al. (2009). We refer to our solar abundances when determining the stellar abundances ratios, [X/H] and [X/Fe].    

\begin{table*}
\centering
\begin{minipage}{165mm}
\caption{Basic photometric and spectroscopic atmospheric parameters for the stars in each association.}
\label{tab3}
\begin{tabular}{lccccccccc}  \hline
\multicolumn{1}{l}{Star} & \multicolumn{3}{c}{T$^{\rm phot}_{\rm eff}$ (K)}& \multicolumn{1}{c}{T$^{\rm spec}_{\rm eff}$}& 
\multicolumn{1}{c}{$\log g_{\rm spec}$}& \multicolumn{1}{c}{$\xi_{\rm spec}$}& \multicolumn{1}{c}{$v$~sin$i$ (km sec$^{-1}$)} 
& \multicolumn{1}{c}{$v$~sin$i$ (km sec$^{-1}$)} & \multicolumn{1}{c}{ Ref$^{a}$ } \\  \cline{2-4} 
\multicolumn{1}{c}{} & \multicolumn{1}{c}{(B-V)}& (V-K$_{\rm s}$)& (J-K$_{\rm s}$)& \multicolumn{1}{c}{(K)} & \multicolumn{1}{c}{(cm s$^{-2}$)}& 
\multicolumn{1}{c}{(km sec$^{-1}$)}& \multicolumn{1}{c}{(This work)} & \multicolumn{1}{c}{(Literature)} & \multicolumn{1}{c}{ } \\
\hline

BD-20 2977 & 5543 & 5399 & 5420 & 5500 & 4.40 & 1.51 & 10.0$\pm$1.5 & 10.1 & [1] \\
HD 61005   & 5486 & 5486 & 5412 & 5550 & 4.55 & 1.50 &  9.8$\pm$1.5 &  8.2 & [1] \\
HD 103742  & 5783 & 5892 & 5724 & 5800 & 4.50 & 1.32 &  5.8$\pm$1.5 &  5.0$\pm$2.0 & [2] \\
HD 103743  & 5691 & 5802 & 5728 & 5800 & 4.60 & 1.36 &  8.8$\pm$1.5 &  9.0$\pm$2.0 & [2] \\
 HD 166    & 5458 & 5420 & 5527 & 5500 & 4.50 & 1.16 &  4.5$\pm$1.5 &  4.0 & [3] \\
 HD 10008  & 5322 & 5313 & 5313 & 5300 & 4.50 & 0.94 &  2.5$\pm$1.5 &  1.7 & [4] \\
 HD 17925  & 5168 & 5227 & 5008 & 5150 & 4.50 & 1.30 &  5.0$\pm$1.5 &  4.8 & [3] \\
Parenago 2339 &6223&6113 & 6266 & 6050 & 4.25 & 1.93 & 15.0$\pm$1.5 & 15.0 & [5] \\
Parenago 2374 &6151&5608 & 5886 & 5950 & 4.25 & 1.22 &  4.3$\pm$1.5 &  6.0 & [5] \\
HD 113449  & 5119 & 5033 & 5044 & 5150 & 4.30 & 1.05 &  5.8$\pm$1.5 &  4.8 & [4] \\
HD 152555  & 5912 & 5904 & 5886 & 6050 & 4.40 & 1.91 & 16.5$\pm$1.5 & 10.0 & [6] \\

\hline
\end{tabular}
$^{a}$Reference $v$~sin$i$ estimates from the literature: [1] De Silva et al. (2013); [2] D'Orazi et al. (2012); [3] Mishenina et al. (2012); [4] Jenkins et al. (2011); [5] Cunha et al. (1995); [6] White et al. (2007)
\end{minipage}
\end{table*}

\subsection{Determination of atmospheric parameters} \label{Dap}

As the EW of a spectral line is affected by the physical conditions and number density of absorbers in the stellar atmosphere, it is necessary to predetermine the atmospheric parameters to estimate the stellar abundances. We derived the star's photometric effective temperatures by substituting the available (B-V), (V-K$_{\rm s}$) and (J-K$_{\rm s}$) colors into the metallicity-dependent color$-$temperature calibrations of Casagrande et al. (2010). Stars sampled for the present analysis are within 100 pc of the Sun  (see Table \ref{tab1}) thereby making the interstellar extinction unimportant for deriving the star's effective temperature, i.e., the observed colors are not corrected for reddening.

We improved our photometric estimate of effective temperatures using spectroscopy adopting an initial value of surface gravity almost solar, i.e., log~$g = 4.44$. The line list (Table \ref{EWmeasurement}) and model atmospheres were then used as inputs to the 2010 version of LTE line analysis and spectrum synthesis code {\scs \bf MOOG}\footnote{{\scs \bf MOOG} was developed and updated by Chris Sneden and originally described in Sneden (1973)} for an abundance analysis. The one-dimensional, line-blanketed plane-parallel atmospheres constructed assuming LTE, hydrostatic equilibrium and flux (radiative plus convective) conservation with the overshooting option switched off were interpolated linearly from the ATLAS9 model atmosphere grid of Castelli \& Kurucz (2003). 

We performed a differential abundance analysis relative to the Sun by running the {\it abfind} driver in {\scs \bf MOOG} using a model with photometric estimates of atmospheric parameters and the measured EWs. The individual line abundances were derived by force-fitting the model generated EWs to the observed ones (see \S \ref{line_selection}) while satisfying the following three constraints simultaneously: First, the microturbulence, $\xi_{t}$, assumed to be isotropic and depth independent is determined by imposing the condition that the Fe abundance from Fe {\scs I} lines be independent of a line's EW. Second, the effective temperature, $T_{\rm eff}$, is estimated by ensuring that the Fe abundance from Fe {\scs I} lines (as they cover a good range in LEP $\sim$ 0.0 to 5.0 eV) showed no trend with line's LEP (excitation equilibrium). $T_{\rm eff}$ is varied in steps of $\pm$25 K from it's photometric value until the slope between log\,$\epsilon$(Fe {\scs I}) and line's LEP was $<$ 0.005. Third, the surface gravity, log~$g$, is adjusted in steps of $\pm$0.05 until the Fe abundance from Fe {\scs I} lines agreed within 0.02 dex to that of Fe {\scs II} lines (i.e., maintaining ionization equilibrium between the neutral and ionized species) for the derived $T_{\rm eff}$ and $\xi_{t}$. As all these atmospheric parameters are interdependent, several iterations are needed to choose the best model from the grid.

\begin{figure}
\begin{center}
\includegraphics[trim=4.6cm 5.8cm 2.8cm 7.4cm, clip=true,height=0.32\textheight,angle=-90]{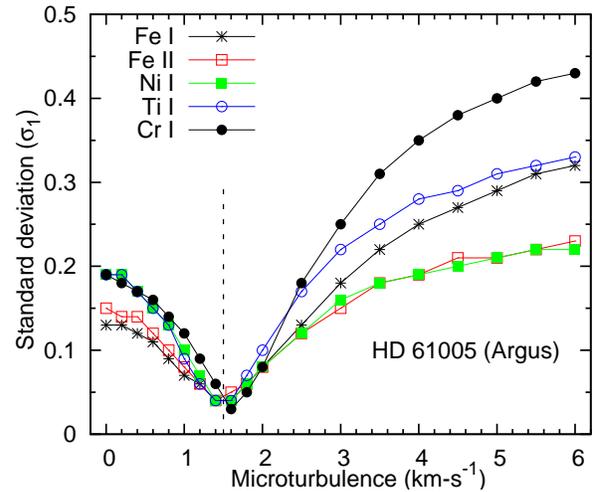}
\caption[]{The standard deviation about the mean abundance derived from the suite of Fe {\scs I}, Fe {\scs II}, Ni {\scs I}, Ti {\scs I}, and Cr {\scs I} lines as a function of $\xi_{t}$ for a model with $T_{\rm eff}$= 5550 K, log~$g$=4.55 cm s$^{-2}$ and [Fe/H]= 0.0 dex. The dotted vertical line refers to $\xi_{t}$ of 1.5 km s$^{-1}$ derived using Fe {\scs I} lines. }
\label{micro_turb}
\end{center}
\end{figure}

The uncertainty in $T_{\rm eff}$ was estimated to be the temperature difference from the model value that would introduce spurious trends in log\,$\epsilon$(Fe {\scs I}) versus LEPs corresponding to the Fe abundance deviating larger than the $\pm$1$\sigma$ standard deviation of the determined Fe abundance. This condition is also verified with the lines of other species like Ti {\scs I} and Ni {\scs I}.  Similarly, the uncertainty in $\xi_{t}$ would be the the difference in $\xi_{t}$ that would introduce spurious trends in log\,$\epsilon$(Fe {\scs I}) versus reduced EWs which equals the variation of Fe abundance larger than its $\pm$1$\sigma$ value. A check on the $\xi_{t}$ is provided by lines of Fe {\scs I}, Fe {\scs II}, Ni {\scs I}, Ti {\scs I}, and Cr {\scs I} species as illustrated in the Figure \ref{micro_turb} for HD 61005. For a chosen model, we compute the dispersion in mean abundances from the suite of Fe {\scs I}, Fe {\scs II}, Ni {\scs I}, Ti {\scs I}, and Cr {\scs I} lines for $\xi_{t}$ from 0.0 to 6.0 km s$^{-1}$. It is clear that the minimum value of dispersion for all species is in the range 1.4-1.6 km s$^{-1}$. Thus, the $\xi_{t}$= 1.5$\pm$0.1 km s$^{-1}$ derived for Fe {\scs I} lines also satisfies the other species. The error in log~$g$ is assumed to be the difference in surface gravities that produce a difference larger than $\pm$1$\sigma$ value in the Fe abundances derived from Fe {\scs I} and Fe {\scs II} lines. A check on this condition is also performed by the abundances derived from the suite of Sc, Ti, and Cr species as they provide both neutral and ionized lines. The typical errors adopted in our analysis are $\pm$75 K in $T_{\rm eff}$, 0.10 cm s$^{-2}$ in log~$g$ and 0.10 km s$^{-1}$ in $\xi_{t}$. The derived stellar parameters for program stars in each of the association are given in Table \ref{tab3}. 

\subsection{Abundances and error estimation} \label{Abuerror}

\begin{figure*}
\begin{center}
\includegraphics[trim=0.1cm 6.8cm 0.1cm 4.0cm, clip=true,height=0.26\textheight,width=0.7\textheight]{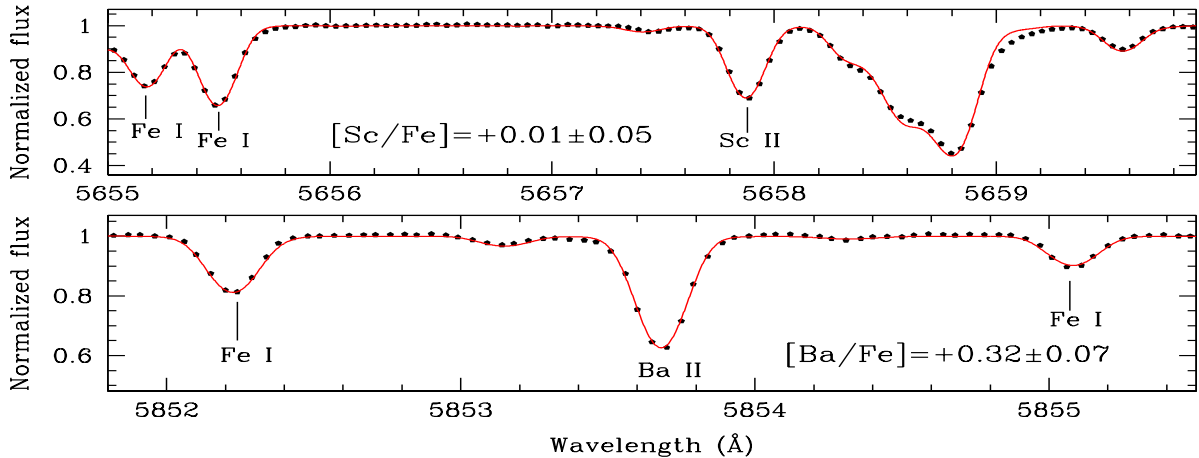}
\caption[]{Comparison of synthetic (red lines) and observed (black dots) spectra for the star HD 103742 (Carina-Near) for regions surrounding Sc {\scs II} and Ba {\scs II} lines. In each panel we show the best fit abundance value with corresponding error due to uncertainties in the best fit value and in atmospheric parameters. }
\label{synth}
\end{center}
\end{figure*}

The abundance analysis was completed by running the {\it abfind} and {\it synth} drivers of {\scs \bf MOOG} for other species in the line list with the atmospheric parameters determined from the Fe lines. In most cases, the abundances are derived from the measured EWs but a few lines were analysed with synthetic spectra. 

\begin{table}
% \centering
{\fontsize{8}{8}\selectfont
\begin{minipage}{90mm}
\caption{Sensitivity of derived abundances to uncertainties in stellar parameters for the star HD 166 (Hercules-Lyra) $T_{\rm eff}$= 5500 K, $\log{g}$= 4.50 cm s$^{-2}$,and $\xi_{t}$= 1.16 km s$^{-1}$ in our sample. Each number refer to the difference between the abundances obtained with and without varying each stellar parameter separately, while keeping the other two unchanged.  }
\label{sensitivity}
\begin{tabular}{lcccc}   \hline
\multicolumn{1}{l}{ }&\multicolumn{1}{l}{T$_{\rm eff}\pm$75}&\multicolumn{1}{l}{$\log~{g}\pm$ 0.10} &
\multicolumn{1}{l}{$\xi_{t}\pm$ 0.10} & \multicolumn{1}{l}{ } \\ \cline{2-5}
\multicolumn{1}{l}{Species}&\multicolumn{1}{c}{$\sigma_{T_{\rm eff}}$}&\multicolumn{1}{c}{$\sigma_{log{g}}$} & 
\multicolumn{1}{c}{$\sigma_{\xi_{t}}$} &\multicolumn{1}{c}{$\sigma_{2}$}  \\ \hline

Li {\scs I} & $+0.08/-0.08$ &$~0.00/~0.00$  &$~0.00/~0.00$   & 0.08  \\
O {\scs I}  & $-0.03/+0.08$ &$ 0.00/-0.01$  &$ 0.00/+0.01$   & 0.05  \\
Na {\scs I} & $+0.05/-0.04$ &$-0.02/+0.01$  &$-0.01/+0.01$   & 0.05  \\
Mg {\scs I} & $+0.03/-0.03$ &$-0.02/+0.01$  &$-0.01/~0.00$   & 0.03  \\
Al {\scs I} & $+0.04/-0.02$ &$-0.01/+0.02$  &$ 0.00/+0.01$   & 0.03  \\
Si {\scs I} & $+0.01/+0.02$ &$+0.02/ 0.00$  &$ 0.00/+0.01$   & 0.02  \\
Ca {\scs I} & $+0.06/-0.05$ &$-0.02/+0.01$  &$-0.02/+0.02$   & 0.06  \\
Sc {\scs I} & $+0.08/-0.07$ &$ 0.00/-0.01$  &$ 0.00/+0.01$   & 0.07  \\
Sc {\scs II}& $~0.00/+0.01$ &$+0.05/-0.05$  &$-0.02/+0.01$   & 0.05  \\ 
Ti {\scs I} & $+0.08/-0.09$ &$-0.01/-0.01$  &$-0.03/+0.02$   & 0.09  \\
Ti {\scs II}& $~0.00/+0.02$ &$+0.06/-0.04$  &$-0.02/+0.02$   & 0.05  \\
V  {\scs I} & $+0.08/-0.09$ &$~0.00/-0.01$  &$-0.01/+0.01$   & 0.08  \\
Cr {\scs I} & $+0.07/-0.06$ &$-0.01/~0.00$  &$-0.02/+0.02$   & 0.06  \\ 
Cr {\scs II}& $-0.01/+0.05$ &$+0.07/-0.03$  &$-0.01/+0.01$   & 0.06  \\ 
Mn {\scs I} & $+0.05/-0.09$ &$-0.01/~0.00$  &$-0.03/+0.03$   & 0.08  \\
Fe {\scs I} & $+0.05/-0.03$ &$~0.00/~0.00$  &$-0.02/+0.03$   & 0.05  \\
Fe {\scs II}& $-0.02/+0.05$ &$+0.07/-0.04$  &$-0.03/+0.02$   & 0.07  \\ 
Co {\scs I} & $+0.04/-0.05$ &$+0.01/-0.02$  &$-0.01/+0.01$   & 0.05  \\ 
Ni {\scs I} & $+0.03/-0.02$ &$+0.02/-0.02$  &$-0.02/+0.02$   & 0.04  \\ 
% Cu {\scs I} & $+0.03/-0.02$ &$+0.01/-0.02$  &$-0.03/+0.02$   & 0.04  \\
Zn {\scs I} & $-0.01/+0.03$ &$+0.04/-0.02$  &$-0.03/+0.02$   & 0.04  \\
%Rb I& $+0.07/-0.08$ &$-0.01/ 0.00$  &$-0.01/ 0.00$   & 0.04  \\
Y {\scs II} &$+0.01/~0.00$  &$+0.06/-0.04$  &$-0.02/+0.02$   & 0.05  \\
Zr {\scs I} &$+0.10/-0.09$  &$~0.00/-0.01$  &$~0.00/+0.01$   & 0.09  \\
Zr {\scs II}&$ 0.00/+0.01$  &$+0.05/-0.04$  &$-0.01/~0.00$   & 0.04  \\
Ba {\scs II}&$+0.02/-0.01$  &$+0.05/-0.04$  &$-0.06/+0.06$   & 0.07  \\
La {\scs II}&$+0.02/-0.01$  &$+0.05/-0.05$  &$-0.01/~0.00$   & 0.05  \\
Ce {\scs II}&$+0.02/-0.01$  &$+0.05/-0.05$  &$-0.01/+0.01$   & 0.05  \\
Nd {\scs II}&$+0.02/-0.01$  &$+0.05/-0.04$  &$-0.01/+0.02$   & 0.05  \\
Sm {\scs II}&$+0.03/-0.01$  &$+0.05/-0.04$  &$~0.00/+0.01$   & 0.05  \\
Eu {\scs II}&$~0.00/+0.01$  &$+0.05/-0.04$  &$-0.01/~0.00$   & 0.04  \\

\hline
\end{tabular}
\end{minipage} } \vspace{-0.4cm}
\end{table} 

Synthetic spectra were computed for species affected by hyperfine structure (hfs) and isotopic splitting and/or affected by blends and matched to the observed spectra by adjustment of abundances. The features analysed by spectrum synthesis included: Li {\scs I} (6707 \AA), (Sc {\scs II} (5657 \AA), Mn {\scs I} (6013 \AA\ \& 6021 \AA), Zn {\scs I} (4810 \AA), Ba {\scs II} (5853 \AA), Sm {\scs II} (4577 \AA), and Eu {\scs II} (6645 \AA). The adopted hfs data and isotopic ratios are same as those listed in Reddy et al. (2012). 

Our linelists have been tested extensively to reproduce the solar and Arcturus spectra before applying them to the stellar spectra using the \textit{synth} driver in {\scs \bf MOOG}. The synthetic spectra were smoothed with a Gaussian profile with a width that represents the broadening due to the instrumental profile and the rotation of the star. The derived rotational velocities ($v$~sin$i$) of our sample stars are listed in Table \ref{tab3}. Good agreement is found between ours and previous determinations of $v$~sin$i$. Synthetic spectra fits to the observed spectrum of HD 103742 are shown in Figure \ref{synth} for regions surrounding Sc {\scs II} and Ba {\scs II} lines. 

In Tables \ref{abu_Argus}$-$\ref{abu_Hercules} we provide abundance results for individual stars averaged over all available lines of given species, relative to solar abundances derived from the adopted $gf$-values. The tables give the average [Fe/H] and [X/Fe] for all elements, and standard deviation along with the number of lines used in calculating the abundance of that element in a parentheses. Abundances calculated by synthesis are presented in bold typeface. Examination of the Tables \ref{abu_Argus}$-$\ref{abu_Hercules} show that stars in a given association have identical [X/Fe] for all elements to within the standard deviations computed for individual stars. From the spread in the abundances for the stars of a given association we obtain the standard deviation $\sigma_1$ in the Tables \ref{abu_Argus}$-$\ref{abu_Hercules} in the column headed `average'. 

\begin{figure*}
\begin{center}
\includegraphics[trim=0.1cm 9.7cm 0.8cm 4cm,clip=true,width=0.9\textwidth]{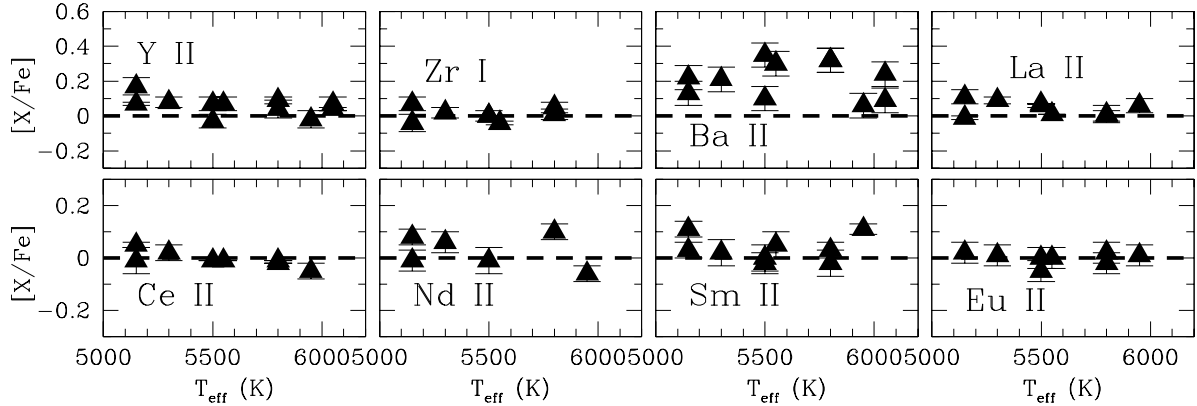}
\caption[]{[X/Fe] ratios for the heavy elements as a function of the effective temperatures of the stars, showing no effects of over-ionization/excitation.}
\label{heavyvsteff}
\end{center}
\end{figure*}

We evaluated the impact of stellar parameters (T$_{\rm eff}$, log~$g$, $\xi_{t}$) on the derived abundances by varying each parameter separately by an amount equal to its uncertainty, while keeping the other two unchanged. The changes in the abundance caused by varying T$_{\rm eff}$ by $\pm$ 75 K, log~$g$ by 0.1 cm s$^{-2}$ and $\xi_{t}$ by 0.1 km s$^{-1}$ with respect to the chosen model atmosphere are summarized in Table \ref{sensitivity}. The quadratic sum of all the three contributors are presented in the column headed $\sigma_{2}$. The total error $\sigma_{tot}$ for each of the element is the quadratic sum of $\sigma_{1}$ and $\sigma_{2}$. The final mean chemical composition of each association along with the $\sigma_{tot}$ from this study are presented in Table \ref{mean_abundance}. We have verified that our abundance estimates for all association members are free of over-ionization and over-excitation effects as reported in De Silva et al. (2013) for their sample of dwarf stars. The Figure \ref{heavyvsteff} shows the typical result for the [X/Fe] ratios for the heavy elements versus T$_{\rm eff}$ for our sample of FGK dwarfs.
% It is evident from the figure that our sample of FGK dwarfs does not show over-ionization/excitation effects. 

\begin{table*}
\caption{Mean abundance ratios, [X/Fe], for elements from Na to Eu for the Argus, Carina-Near, Hercules-Lyra, Orion and Subgroup B4 associations. Abundances calculated by synthesis are presented in bold typeface.} 
\label{mean_abundance}
\begin{tabular}{lccccc}   \hline
\multicolumn{1}{c}{Species}  & \multicolumn{1}{c}{ Argus} & \multicolumn{1}{c}{ Carina-Near } &  
\multicolumn{1}{c}{Hercules-Lyra} & \multicolumn{1}{c}{Orion} & \multicolumn{1}{c}{Subgroup B4}   \\ \hline
                            
% log\,$\epsilon$(Li) & $\bf+2.32\pm0.06$ & $\bf+1.90\pm0.06$ & $\bf+1.30\pm0.05$ & $\bf+1.68\pm0.06$ & $\bf+1.85\pm0.06$  \\ 
$[$O I/Fe$]$   & $+0.09\pm0.04$ & $+0.11\pm0.04$ & $+0.05\pm0.03$ & $+0.22\pm0.05$ & $+0.09\pm0.04$ \\
$[$Na I/Fe$]$  & $+0.07\pm0.05$ & $-0.05\pm0.04$ & $-0.03\pm0.03$ & $+0.01\pm0.05$ & $-0.06\pm0.04$ \\
$[$Mg I/Fe$]$  & $-0.08\pm0.03$ & $-0.05\pm0.03$ & $-0.07\pm0.02$ & $+0.01\pm0.02$ & $-0.03\pm0.03$ \\
$[$Al I/Fe$]$  & $-0.02\pm0.03$ & $-0.09\pm0.04$ & $-0.10\pm0.02$ & $+0.02\pm0.03$ & $-0.08\pm0.03$ \\
$[$Si I/Fe$]$  & $ 0.00\pm0.02$ & $+0.03\pm0.03$ & $+0.03\pm0.02$ & $+0.07\pm0.03$ & $+0.05\pm0.02$ \\
$[$Ca I/Fe$]$  & $+0.06\pm0.05$ & $+0.04\pm0.05$ & $+0.02\pm0.04$ & $+0.03\pm0.05$ & $+0.03\pm0.05$ \\
$[$Sc I/Fe$]$  & $-0.01\pm0.05$ & $-0.02\pm0.06$ & $-0.04\pm0.05$ &   $\ldots$     &    $\ldots$     \\
% $[$Sc II/Fe$]$ & $-0.01\pm0.04$ & $-0.01\pm0.05$ & $-0.01\pm0.04$ & $+0.10\pm0.07$ & $-0.06\pm0.05$ \\
$[$Sc II/Fe$]$ &$\bf+0.05\pm0.04$ &$\bf+0.06\pm0.04$ &$\bf+0.01\pm0.03$ &$\bf+0.01\pm0.04$ & $\bf+0.03\pm0.04$ \\
$[$Ti I/Fe$]$  & $ 0.00\pm0.07$ & $+0.02\pm0.07$ & $+0.01\pm0.05$ & $+0.01\pm0.07$ & $ 0.00\pm0.07$ \\
$[$Ti II/Fe$]$ & $+0.02\pm0.04$ & $-0.05\pm0.05$ & $-0.04\pm0.03$ & $-0.01\pm0.04$ & $-0.06\pm0.04$ \\
$[$V I/Fe$]$   & $+0.04\pm0.06$ & $+0.01\pm0.06$ & $+0.04\pm0.05$ & $+0.04\pm0.06$ & $+0.03\pm0.06$ \\
$[$Cr I/Fe$]$  & $+0.05\pm0.05$ & $+0.03\pm0.05$ & $+0.03\pm0.04$ & $-0.07\pm0.05$ & $-0.05\pm0.05$ \\
$[$Cr II/Fe$]$ & $+0.02\pm0.05$ & $+0.01\pm0.05$ & $+0.07\pm0.04$ & $-0.01\pm0.05$ & $-0.01\pm0.05$ \\
$[$Mn I/Fe$]$ &$\bf+0.06\pm0.06$ &$\bf+0.02\pm0.06$ &$\bf+0.06\pm0.05$ &$\bf-0.02\pm0.06$ &$\bf-0.04\pm0.06$   \\
$[$Fe I/H$]$   & $-0.01\pm0.05$ & $ 0.00\pm0.04$ & $+0.02\pm0.03$ & $-0.10\pm0.05$ & $-0.06\pm0.05$ \\
$[$Fe II/H$]$  & $ 0.00\pm0.05$ & $ 0.00\pm0.06$ & $+0.03\pm0.04$ & $-0.10\pm0.06$ & $-0.05\pm0.06$ \\
$[$Co I/Fe$]$  & $-0.12\pm0.04$ & $-0.10\pm0.04$ & $-0.05\pm0.03$ & $-0.05\pm0.04$ & $-0.07\pm0.04$ \\
$[$Ni I/Fe$]$  & $-0.07\pm0.04$ & $-0.04\pm0.04$ & $-0.04\pm0.03$ & $-0.06\pm0.04$ & $-0.05\pm0.04$ \\
% $[$Cu I/Fe$]$  & $-0.06\pm0.03$ & $-0.12\pm0.04$ & $-0.06\pm0.05$ & $-0.01\pm0.06$ & $-0.04\pm0.03$ \\
% $[$Cu I/Fe$]$  &$\bf-0.27\pm0.03$ &$\bf-0.30\pm0.03$ &$\bf-0.14\pm0.03$ &$\bf-0.09\pm0.03$ &$\bf-0.17\pm0.03$ \\
$[$Zn I/Fe$]$  &$\bf-0.07\pm0.03$ &$\bf-0.10\pm0.03$ &$\bf-0.09\pm0.03$ &$\bf+0.06\pm0.03$ &$\bf-0.11\pm0.03$ \\
% $[$Rb I/Fe$]$  &  $\bf+0.05$   &  $\bf+0.15$   &  $\bf-0.10$ &  $\bf+0.17$ &  $\bf-0.10$  \\
$[$Y II/Fe$]$  & $+0.07\pm0.05$ & $+0.06\pm0.05$ & $+0.07\pm0.03$ & $+0.02\pm0.05$ & $+0.07\pm0.04$ \\
$[$Zr I/Fe$]$  & $-0.02\pm0.06$ & $+0.03\pm0.07$ & $+0.03\pm0.05$ &  $\ldots$      & $-0.04\pm0.08$ \\
% $[$Zr II/Fe$]$ &   $\ldots$   &   $\ldots$     & $-0.01\pm0.04$   &   $\ldots$     &   $\ldots$     \\
$[$Ba II/Fe$]$ &$\bf+0.32\pm0.05$ &$\bf+0.32\pm0.05$ &$\bf+0.15\pm0.04$ &$\bf+0.07\pm0.05$ &$\bf+0.23\pm0.05$ \\
$[$La II/Fe$]$ & $+0.04\pm0.04$ & $ 0.00\pm0.05$ & $+0.09\pm0.03$ & $+0.06\pm0.05$ & $-0.01\pm0.04$ \\
$[$Ce II/Fe$]$ & $-0.01\pm0.04$ & $-0.01\pm0.04$ & $ 0.00\pm0.03$ & $-0.05\pm0.05$ & $+0.05\pm0.04$ \\
$[$Nd II/Fe$]$ &   $\ldots$     & $+0.10\pm0.05$ & $+0.01\pm0.04$ & $-0.06\pm0.05$ & $+0.08\pm0.05$ \\
$[$Sm II/Fe$]$ &$\bf+0.02\pm0.04$ &$ 0.00\pm0.03$ &$+0.01\pm0.02$ &$+0.11\pm0.02$ &$+0.11\pm0.03$ \\
$[$Eu II/Fe$]$ &$\bf 0.00\pm0.03$ &$\bf 0.00\pm0.03$ &$\bf-0.01\pm0.03$ &$\bf+0.01\pm0.03$ &$\ldots$  \\

\hline
\end{tabular}
\end{table*}

\section{Results}

\subsection{Summary}

The overall impression of Table \ref{mean_abundance} is that these five local associations have a solar composition with [Fe/H] and [X/Fe] within zero to within the uncertainty $\sigma_{tot}$ for essentially every element X. Exceptions include Li which is highly depleted in the Sun, and Ba which, as we noted in the Introduction, has been reported as highly overabundant in young open clusters and slightly overabundant in local associations. Lithium and barium are discussed below in separate subsections. 

That local gas, young stars, clusters and associations have essentially a solar composition is not a new result. This result represents a puzzle for the simplest of models Galactic Chemical Evolution: Why after 4.5 Gyr of stellar nucleosynthesis does the region around the Sun have a solar composition and not one enriched in products of stellar nucleosynthesis? For example, this puzzle in connection with star-forming regions is   discussed by James et al. (2006) and Santos et al. (2008), early B-type stars by Lyubimkov et al. (2005) and Nieva \& Przybilla (2012), H {\scs II} regions by Smartt \& Rolleston (1997) and Rood et al. (2007), and young clusters by D'Orazi \& Randich (2009b) and Biazzo et al. (2011) in the solar neighbourhood.

\subsection{Lithium abundance}

\begin{figure}
\begin{center}
\includegraphics[trim=2.4cm 2.6cm 0.8cm 0.1cm,clip=true,width=0.4\textwidth,angle=-90]{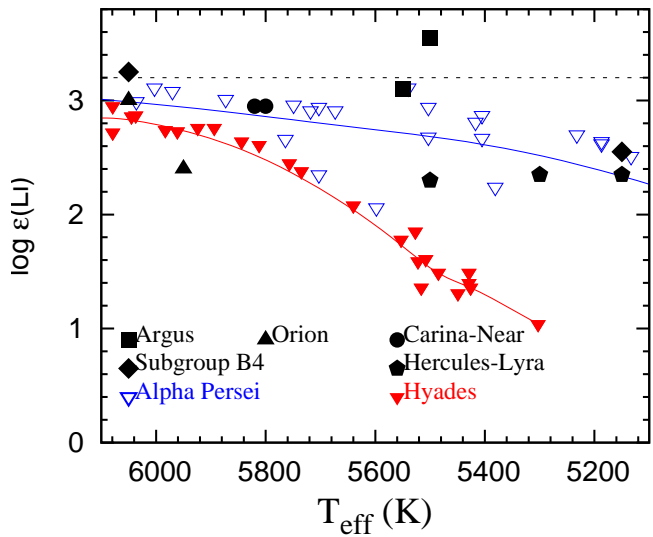}
\caption[]{Lithium abundance versus effective temperature for associations, $\alpha$ Per (Balachandran et al. 2013) and the Hyades (Takeda et al. 2013). The lines represent the smooth curves passing through the lithium abundance of $\alpha$ Persei and the Hyades members.}
\label{lithium}
\end{center}
\end{figure}

In young clusters such as the Pleiades (age = 100 Myr) and $\alpha$ Per (age = 50 Myr), it has been shown by Soderblom et al. (1993) and  Balachandran, Mallik \& Lambert (2011) for the Pleiades and $\alpha$ Per, respectively, that the Li abundance in stars approaching the main sequence declines from log\,$\epsilon$(Li) $\simeq$ 3.3 at T$_{\rm eff} \simeq 7000$ K to about log\,$\epsilon$(Li) $\simeq$ 2.3 at T$_{\rm eff} \simeq 5100$ K, the lowest temperature in our sample. In this run of Li with $T_{\rm eff}$, real scatter at a given effective temperature is found for (see Figure \ref{lithium}) $\alpha$ Per and possibly also for the Pleiades. Given that we have only two or three stars from a given association, we compare in Figure \ref{lithium} an association's Li abundance with the abundances reported for $\alpha$ Per. 

In order of increasing age, the associations are Orion (3 Myr), Argus (30 Myr), $\alpha$ Per (50 Myr), Subgroup B4  (80 Myr),  Carina-Near (200 Myr) and Hercules-Lyra (200 Myr). For the Orion association, Parenago 2339 at T$_{\rm eff} = 6050$ K falls on the $\alpha$ Per trend but Parenago 2374 at T$_{\rm eff} = 5950$ K is depleted by about 0.6 dex relative to its $\alpha$ Per counterpart. Lithium in this pair of stars was measured by Cunha et al. (1995) who obtained slightly lower Li abundances but the differences are due to the 50$-$100 K lower effective temperatures adopted by Cunha et al. For a discussion of Li in these and other Orion F$-$G members see Cunha et al. (1995). The pair of Argus stars have similar effective temperatures and straddle the assumed initial Li abundance and thus skirt the upper envelope of the distribution given by $\alpha$ Per. The stars from Subgroup B4 -- one warm and one cool -- fall within the $\alpha$ Per trend. 

Carina-Near, although older than $\alpha$ Per by a factor of perhaps four, has according to HD 103742 and HD 103743, the same slightly depleted  Li abundance. One might note that in the Hyades with an age of 660 Myr Li at the same effective temperature is log\,$\epsilon$(Li) $\simeq$ 2.6 dex (Takeda et al. 2013). The lithium abundances of the three stars from Hercules-Lyra are not distinguishable from their $\alpha$ Per counterparts. Their Li abundances are 1 dex or more greater than those in the 660 Myr Hyades cluster. Hercules-Lyra and to a weaker extent Carina-Near suggest that depletion occurring in Hyades and similar older clusters may  occur at a faster rate after about 200 Myr. This suggestion is one reason why a more thorough study of Li in these associations would be of great interest. (For a more precise comparison of stars in associations and clusters of different ages, one should compare stars of the same mass rather the same effective temperature.)

\subsection{Argus}
The chemical abundances for some of the elements of Argus group members were determined previously by three different studies: Desidera et al. (2011), VA09 and De Silva et al. (2013). Desidera et al. (2011) have analysed the star HD 61005 and derived the following values: $T_{\rm eff}$=5500 K, log~$g$=4.55 cm s$^{-2}$, $\xi_{t}$=1.0 km s$^{-1}$, [Fe/H]$=0.01\pm0.04$, [Na/Fe]$=-0.03\pm0.05$, [Si/Fe]$=0.00\pm0.04$, [Ca/Fe]$=0.00\pm0.04$, [Ti/Fe]$=0.02\pm0.06$, [Ni/Fe]$=0.00\pm0.06$, and [Ba/Fe]$=0.63\pm0.06$. The same star was included in our analysis for which our derivation of $T_{\rm eff}$ and log~$g$ matches excellently while $\xi_{t}$ is 0.5 km s$^{-1}$ higher than their value (Table \ref{tab3}). Our abundance estimates [X/Fe] for this star (Table \ref{abu_Argus}) agree within 0.08 dex for all elements in common with the exception of Ba for which our value of [Ba/Fe] is lower than theirs by 0.33 dex. As the strong Ba feature employed in the abundance analyses is sensitive to the adopted microturbulence, the relatively low Ba content obtained in our study is not surprising. Had we adopt their value of microturbulence, our Ba abundance will be increased by 0.3 dex sufficient to provide good agreement between the two analyses.

VA09 presented abundances for seven members, obtaining mean values of [Fe/H]$=-0.03\pm0.05$, [Ni/Fe]$=0.02\pm0.04$, and [Si/Fe]$=-0.03\pm0.03$. Comparing these results with our mean abundances of [Fe/H]$=-0.01\pm0.03$, [Ni/Fe]$=-0.07\pm0.03$, and [Si/Fe]$=0.00\pm0.02$ shows that the two studies are in excellent agreement. We have one star, namely BD-20 2977, in common with their study. The difference in effective temperature between our value and theirs is 79 K and gravities agree well within 0.1 dex, while they derived a slightly higher value of microturbulence, i.e., 2.05 km s$^{-1}$ compared with our estimate of 1.51 km s$^{-1}$. Finally, the abundance estimates for [Fe/H], [Ni/Fe], and [Si/Fe] agree within 0.1 dex. 

De Silva et al. (2013) presented abundances for Na, Al, $\alpha$-elements (Mg, Si, Ca, and Ti), iron-peak elements (Cr, Ni, and Fe), and the $s$-process element Ba (and only Ba) for six members. Unfortunately, we do not have stars in common with their work for a direct star-to-star comparison of results, but on the assumption that the intrinsic dispersion in abundances within Argus members is negligible, we may compare results. They found the following mean abundances: [Fe/H]$=-0.05\pm0.05$, [Na/Fe]$=0.05\pm0.04$, [Al/Fe]$=0.03\pm0.02$, [Mg/Fe]$=0.03\pm0.04$, [Si/Fe]$=0.02\pm0.06$, [Ca/Fe]$=0.09\pm0.05$, [Ti/Fe]$=0.03\pm0.04$, [Cr/Fe]$=0.08\pm0.04$, [Ni/Fe]$=0.00\pm0.02$, and [Ba/Fe]$=0.53\pm0.08$. These results agree well within $\pm$0.1 dex for almost all elements in common, with the exception of Ba for which their value of [Ba/Fe] is higher than ours by 0.21. Some of this scatter might be attributed to possible systematic difference in microturbulence; we are able to reproduce their value of [Ba/Fe], if we lower our values of microturbulence by about 0.35 km s$^{-1}$.

In this paper, Argus members are studied comprehensively for many elements relative to Fe including the light $s$-process elements (Y, Zr) and heavy $s$-process elements (Ba, La, Ce, Nd, and Sm) and a $r$-process element Eu. Except for Ba, our abundances for heavy elements are the first to be provided for this association.

\subsection{Carina-Near} \label{car-near}
The chemical composition of the Carina-Near association was measured previously by Biazzo et al. (2012) and D'Orazi et al. (2012) for a sample of 
five and four stars, respectively. Biazzo et al. (2012) published elemental abundances ranging from Li to Zn while D'Orazi et al. (2012) presented abundances for the heavy elements. They concluded that all association members show a solar-like mix of elements (i.e., [X/Fe] $\simeq$ 0), within the small uncertainties, for all elements with the exception Ba, showing an enhancement of almost 0.20$\pm$0.07 dex in [Ba/Fe] ratio. 

We have observed two stars, namely HD 103742, and HD 103743, in common with their work. Our spectroscopic estimates of atmospheric parameters are in good agreement with their work with differences within $\pm$100 K in $T_{\rm eff}$ and $\pm$0.10 cm s$^{-2}$ in log~$g$, while larger discrepancies are noticed for $\xi_{t}$ values. They derived a systematically higher microturbulences for both the stars with $\Delta$$\xi_{t}=$0.28 km s$^{-1}$ (HD 103742) and $\Delta$$\xi_{t}=$0.34 km s$^{-1}$ (HD 103743). 

Using their averaged elemental abundances published for individual stars we calculated the mean abundance for all the elements including the uncertainties in measured EWs and stellar parameters and those values are presented in Table \ref{compare_Carina-near} in comparison with our results. Differences in [X/Fe] between ours and the literature results are $\pm$0.05 dex or smaller for almost all elements: exceptions include Li, Mg, Al, Ti, Zr and Ba but all elements agree well within $\pm$0.1 dex. In this paper, we have extended the abundance estimates to Nd, Sm and Eu which were not considered in previous studies.

\begin{table} \vspace{-0.2cm}
\centering 
\caption{The mean elemental abundances for the two stars, namely HD 103742 and HD 103743, in common between ours and literature analysis. For the literature sample, abundances from Li to Zn are drawn from Biazzo et al. (2012) and the abundances for s-process elements (Y$-$Ce) are taken from D'Orazi et al. (2012). }  \vspace{-0.2cm}
\label{compare_Carina-near}
\begin{tabular}{llc}   \hline
\multicolumn{1}{c}{Species} & \multicolumn{1}{c}{Literature$_{mean}$} & \multicolumn{1}{c}{This work$_{mean}$}  \\ \hline

 log\,$\epsilon$(Li)   & $+2.83\pm\ldots$& $+2.95\pm0.06$ \\
$[$Na {\scs I}/Fe$]$ & $-0.05\pm0.10$ & $-0.05\pm0.04$ \\
$[$Mg {\scs I}/Fe$]$ & $+0.03\pm0.10$ & $-0.06\pm0.03$ \\
$[$Al {\scs I}/Fe$]$ & $-0.01\pm0.10$ & $-0.10\pm0.04$ \\
$[$Si {\scs I}/Fe$]$ & $-0.01\pm0.08$ & $+0.04\pm0.04$ \\
$[$Ca {\scs I}/Fe$]$ & $+0.03\pm0.10$ & $+0.04\pm0.05$ \\
$[$Ti {\scs I}/Fe$]$ & $+0.01\pm0.09$ & $+0.02\pm0.07$ \\
$[$Ti {\scs II}/Fe$]$& $+0.04\pm0.11$ & $-0.05\pm0.05$ \\
$[$Cr {\scs I}/Fe$]$ & $+0.00\pm0.08$ & $+0.04\pm0.05$ \\
$[$Fe {\scs I}/H$]$  & $+0.03\pm0.06$ & $+0.00\pm0.04$ \\
$[$Fe {\scs II}/H$]$ & $+0.03\pm0.06$ & $+0.00\pm0.06$ \\
$[$Ni {\scs I}/Fe$]$ & $-0.07\pm0.10$ & $-0.04\pm0.04$ \\
$[$Zn {\scs I}/Fe$]$ & $-0.08\pm0.08$ & $-0.10\pm0.03$ \\
$[$Y {\scs II}/Fe$]$ & $+0.03\pm0.07$ & $+0.07\pm0.04$ \\
$[$Zr {\scs II}/Fe$]$& $+0.09\pm0.06$ & $+0.03\pm0.07$ \\
$[$Ba {\scs II}/Fe$]$& $+0.24\pm0.12$ & $+0.32\pm0.05$ \\
$[$La {\scs II}/Fe$]$& $+0.05\pm0.06$ & $+0.01\pm0.05$ \\
$[$Ce {\scs II}/Fe$]$& $+0.04\pm0.08$ & $-0.01\pm0.04$ \\

\hline
\end{tabular} \vspace{-0.5cm}
\end{table}

\begin{figure*}
\begin{center}
\includegraphics[trim=0.6cm 11.6cm 0.1cm 4.0cm, clip=true,height=0.18\textheight,width=0.75\textheight]{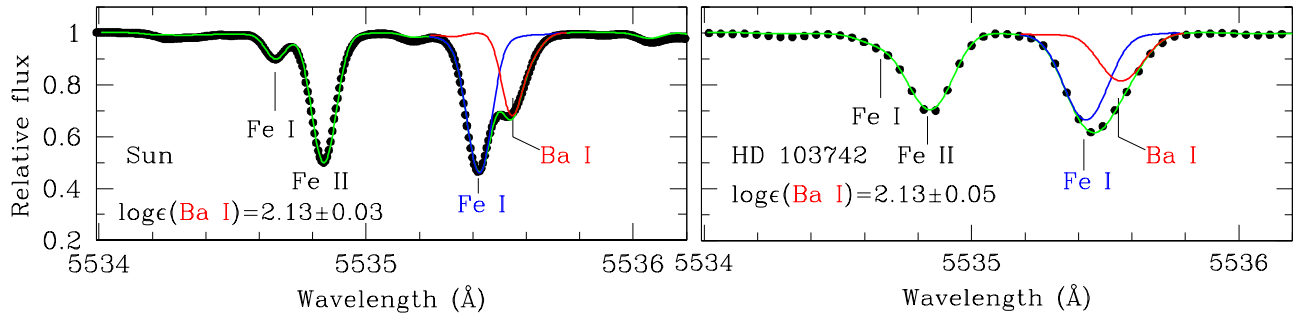}
\caption[]{Comparison of synthetic spectra (green lines) to the observed spectra (black dots) of the Sun (left panel) and to the solar analogue HD 103742 (right panel) in the wavelength region surrounding the Ba {\scs I} feature. The blue and the red lines represent, respectively, the best fit Fe {\scs I} and Ba {\scs I} abundances to the observed spectrum. The target star spectra is convolved with the adopted $v$~sin$i$ in the process of spectrum synthesis. The vertical lines in each panel indicate the central wavelengths of those species. }
\label{synthba1}
\end{center}
\end{figure*}

\subsection{Orion}
The chemical content of Orion association has been explored previously for Li and Fe by Cunha et al. (1995), and for O and Fe by Cunha et al. (1998) for a sample of F and G dwarfs. In this paper, we have done the first detailed chemical abundance study of Orion using two members. We derived a mean iron abundance of [Fe/H]$=-0.10\pm0.03$ dex, which is in agreement with the value of [Fe/H]$=-0.13\pm0.13$ dex reported by Cunha et al. (1998). In particular, for the two stars belong to the sub-group Ic in common with our study, they derived the following mean LTE values: [Fe/H]$=-0.18\pm0.10$, log\,$\epsilon$(Li)=2.8, [O/Fe]$=+0.21\pm0.10$, and [Fe/H]$=-0.10\pm0.09$, log\,$\epsilon$(Li)=2.3, [O/Fe]$=+0.15\pm0.02$, respectively for Parenago 2339 \& 2374. 

For the two stars in common between the analyses the agreement is quiet good for the stellar atmospheric parameters, while a larger difference in $\xi_{t}$ is noticed for Parenago 2339:differences between ours and their values are $\Delta$$T_{\rm eff}$=100 K, $\Delta$log~$g=$ 0.25 cm s$^{-2}$, $\Delta$$\xi_{t}=$ 0.53 km s$^{-1}$, $\Delta$[Fe/H]$=$0.04 dex for Parenago 2339, and $\Delta$$T_{\rm eff}$=50 K, $\Delta$log~$g=$ 0.35 cm s$^{-2}$, $\Delta$$\xi_{t}=$-0.08 km s$^{-1}$, $\Delta$[Fe/H]$=$0.03 dex respectively for Parenago 2374 (see Table \ref{tab3}). A larger value of $\xi_{t}$ derived for Parenago 2339 might be associated with its high rotational velocity. We notice in our study that stars with high rotational velocities are associated with the higher values of microturbulence (see Table \ref{tab3}). Comparing these results with our mean abundances (see Table \ref{abu_Orion}) we can conclude that the two studies are in good agreement. 

\subsection{Hercules-Lyra \& Subgroup B4}
To our knowledge, we have done the first detailed chemical abundance study for these two associations. We found that the two associations share a solar composition, i.e., [X/Fe] $\simeq$ 0, for all elements with the exception of Ba again. Focusing on the heavy elements, the lighter elements Y and Zr as well as the heavier elements La, Ce, Nd and Sm show solar ratios whereas Ba behaves differently yielding slightly higher values. We measured a mean barium abundance of [Ba/Fe]$=0.15\pm0.04$ dex for Hercules-Lyra and [Ba/Fe]$=0.23\pm0.05$ dex for Subgroup B4. This means that the abundance pattern of the two associations is not very different from the other associations analysed in our work.

\section{The barium abundance}

All the associations in our sample have very nearly a solar composition with the exception of Li discussed in Section 5.2  and the puzzling case of Ba. Barium is especially  puzzling in that positive values of [Ba/Fe] of up to $+0.3$ for our sample are not shown by other heavy elements with major contributions expected from the $s$-process which dominates the synthesis of Ba. Our abundances of Ba and other heavy elements closely confirm and extend results for local association obtained by D'Orazi et al. (2012). The upper bound on  Ba is raised when D'Orazi et al.'s (2009) reported Ba abundances ([Ba/Fe] $\simeq +0.6$) for FG dwarfs in very young OCs are considered. 

In this section, the Ba puzzle is explored in three different ways: first, we discuss an assessment of the Ba abundance using for the first time a Ba\,{\sc i} line; second, the sensitivity of the Ba\,{\sc ii} 5853 \AA\ line to microturbulence is examined; third, we comment briefly on abundance variations of the ratio of Ba to other heavy elements which may plausibly be attributed to variations in neutron capture synthesis.

\subsection{The Ba\,{\sc i} line at 5535\AA}

Barium abundance determinations have focussed on one or more Ba\,{\sc ii} lines. Lines of neutral Ba atoms have been neglected on account of the fact that the low ionization potential of the atoms (5.21 eV) ensures that ions Ba$^+$ greatly outnumber Ba atoms in all but the coolest of stars. Here, we explore the possibility that the Ba\,{\sc i} 5535 \AA\ resonance line may be detectable in FG dwarfs and contribute to the resolution of the Ba puzzle. 

The 5535 \AA\ line is blended with a Fe\,{\sc i} line to the blue, thus, spectrum synthesis is used to obtain a Ba abundance. 
At the resolution of our instrument, the Fe\,{\sc i} feature in our solar spectrum is unresolved and entirely blended with the barium feature. So we compared the solar integrated disc spectrum (Kurucz et al. 1984) observed at a resolution of 0.07 \AA\ to the synthetic spectra computed for the solar model to validate our linelist. Our adopted value of log~$gf=-$1.058 to model the solar Fe {\scs I} feature with our reference solar metallicity (i.e., [Fe/H]$=7.52\pm0.4$) is in fair agreement with the experimental $gf$-value provided in the {\scs \bf NIST} database. 

For the Ba {\scs I} line, the number of hfs components and isotopic shifts are drawn from Dammalapati et al. (2009) and the relative strength of each hfs component is calculated using equation 5 in Smitha et al. (2012). As shown in Figure \ref{synthba1}, a reasonably good fit to the solar barium feature (left panel) obtained for the adopted isotopic ratios from Lodders (2003) validates our adopted linelist. The same linelist was applied to the program star spectra. An example of synthetic spectra fit to the observed spectrum of the solar analogue HD 103742 ($T_{\rm eff}$=5800 K, log~$g=$4.5 cm s$^{-2}$ and [Fe/H]=$-$0.02 dex; right panel) is shown in the Figure \ref{synthba1}. The barium abundance from the Ba {\scs I} line [Ba/Fe] $=+0.02\pm0.05$ is appreciably less than the $+0.32\pm0.07$ from the Ba {\scs II} lines.

\begin{figure}
\begin{center}
\includegraphics[trim=0.1cm 9.5cm 3.6cm 4.6cm,clip=true,height=0.19\textheight,width=0.85\textwidth]{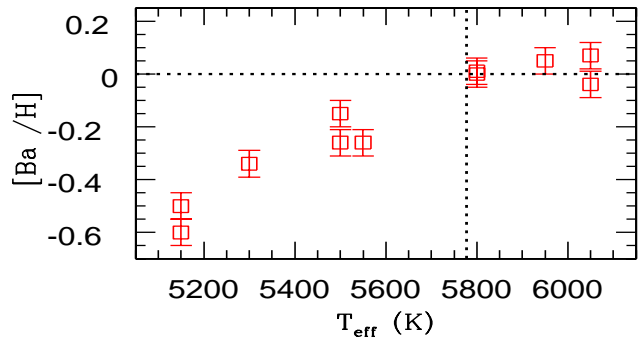}
\caption[]{Our LTE barium abundance from the Ba {\scs I} line as a function of effective temperature. The dotted vertical line indicates the effective temperature of the Sun, i.e., $T_{\rm eff}$=5777 K.}
\label{ba1vsteff}
\end{center}
\end{figure}

\begin{figure}
\begin{center}
\includegraphics[trim=0.6cm 11.6cm 0.1cm 4.2cm, clip=true,height=0.18\textheight,width=0.7\textheight]{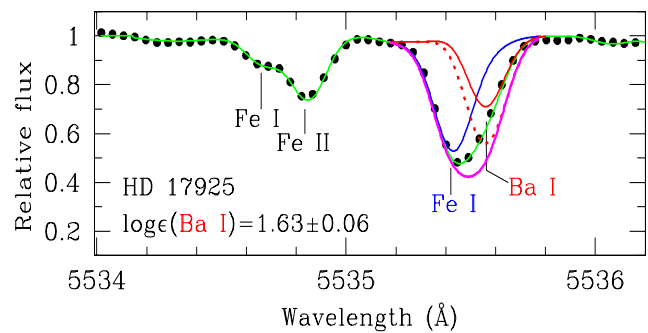}
\caption[]{Same as figure \ref{synthba1}, but for the coolest star HD 17925 ($T_{\rm eff}$= 5150 K) in our study. The broken red line corresponds to the solar value of Ba while the continuous red line represent the best fit Ba abundance of log\,$\epsilon(Ba)=1.63\pm0.06$. The thick magenta line represent the synthetic profile generated for the solar value of Ba at the metallicity of the star.}
\label{synth17925}
\end{center}
\end{figure}

Syntheses of the 5535 \AA\ blend for all of our stars shows that the best-fitting Ba abundance decreases with decreasing effective temperature (Figure \ref{ba1vsteff}). For the coolest star HD 17925 with $T_{\rm eff} = 5150$ K, synthesis calls for [Ba/Fe] $= -0.52$ in sharp disagreement with the [Ba/Fe] $= +0.13$ from the Ba\,{\sc ii} line. Observed and synthetic spectra for HD 17925 in Figure \ref{synth17925} show that the Ba\,{\sc i} line with a solar Ba abundance would make a very striking contribution to the stellar feature but the observed spectrum demands the severe underabundance. The Ba abundance from the Ba\,{\sc i} line is a clear function of $T_{\rm eff}$ with a change from 5100K to 6100 K that far exceeds that expected from the measurement errors: $\Delta T_{\rm eff} = \pm 100$ K leads to $\Delta$ Ba $ = \pm0.10$ dex.   

Our assumption is that the principal cause for the trend in Figure \ref{ba1vsteff} is that non-LTE effects weaken the Ba\,{\sc i} resonance line. Unfortunately, there are to our knowledge  no published non-LTE calculations for the Ba atom; existing calculations such as those by Mashonkina et al. (1999) and Korotin et al. (2015) ignore the Ba atom and consider in detail only the Ba$^+$ ion. Inspection of Figure \ref{ba1vsteff} suggests that the non-LTE effects on the Ba\,{\sc i} resonance line are very small at the effective temperature of the Sun and lead to a steep weakening of the line at lower effective temperatures. The magnitude of the non-LTE effects and their steep dependence on effective temperature are surprising given that calculations for other atoms and ions yield much smaller non-LTE corrections. 

In the absence of accurate non-LTE calculations for the Ba atom, the Ba\,{\sc i} 5535 \AA\ line cannot contribute to the resolution of the Ba puzzle. 

\subsection{Neutron-capture nucleosynthesis}

Results from the discovery phase of the Ba puzzle showed Ba to be overabundant by about $+0.6$ dex in the youngest open clusters (D'Orazi et al. 2009). This enrichment was attributed partly to non-LTE effects, but principally, to the effects of a strong, warm flux from a chromosphere on the structure of the stellar atmosphere. It is a fundamental aspect of $s$-process operation that the higher Ba yield must be accompanied by higher yields of especially La and Ce, also Nd, and Sm but without a measureable change in the Eu abundance (Busso et al. 1999). This conclusion assumes the $r$-process contributions to these elements remain unchanged over time. Contributions of the $s$-process to solar abundances run from 85\% for Ba, 75\% for La, 81\% for Ce, 47\% for Nd, 34\% for Sm and only 3\% for Eu (Burris et al. 2000). Operation of the $s$-process under different conditions will have very little effect on the relative yields of elements Ba to Eu but may change the Ba to Y-Zr ratios depending on the strength of the neutron flux. Y-Zr are considered to be products from AGB stars, the so-called main $s$-process, and from massive stars, the so-called weak $s$-process. 

For the local associations, the Ba puzzle as shown by D'Orazi et al. (2012) and us has shrunk to a Ba overabundance of around $+0.2$ dex but with solar abundances of light (Y, Zr) and other heavy elements (La to Eu). Within the constraints of the $s$-process, it is very difficult to change the Ba/(La-Eu) ratios by $+0.2$ dex either by changes to physical conditions at the $s$-process site or by invocation of new astrophysical sites. This difficulty encourages one to solve the Ba puzzle through alterations to the standard model atmospheres and line formation including the simple device of adapting the microturbulence. 

However, some authors close to the abundance analyses have invoked a long-forgotten neutron-capture process to account for the high Ba abundances. Mishenina et al. (2015) reintroduce the intermediate neutron-capture $i$-process discussed by Cowan \& Rose (1977) in which the neutron density operating the process is approximately 10$^{15}$ cm$^{-3}$. i.e., intermediate between the low densities for the $s$-process and the much higher densities adopted for the $r$-process. However, abundance predictions illustrated by Mishenina at al. (2015) appear not to fit the association heavy element abundances: [Ba/La] $\simeq 1.2$ with [Eu/La] $\simeq -0.5$ are predicted but with a single exception our results are [Ba/La] $\sim 0.2$ with [Eu/La] near $0.00$.

\subsection{Sensitivity to microturbulence}

In seeking a solution to the Ba puzzle, one  cannot but help focus on the fact that the selected Ba\,{\sc ii} 5853 \AA\ line is often the strongest line in a star's linelist. The typical equivalent width is such that, in classical terminology, the line is near or on the the flat part of the curve of growth and, thus, the derived abundance is particularly sensitive to the adopted value of the microturbulence. Inspection of Table \ref{sensitivity} shows that among the considered elements it is the Ba abundance from the Ba\,{\sc ii} which line has the greatest sensitivity to the uncertainty in the microturbulence. 

Table \ref{mean_abundance} shows that [Ba/Fe] runs from $+0.07$ to $+0.32$ for our associations but [X/Fe] is close to zero for other heavy elements and notably those -- La, Ce, Nd, and Sm -- whose abundance is expected to be dominated by the $s$-process which controls the Ba abundance. The positive difference between [Ba/H] and [La/H] et al. is the heart of the Ba puzzle for the associations. This puzzle is more serious  for young OCs in that, 
according to D'Orazi et al. (2009), [Ba/H] for the OCs is approximately  $+0.6$ and [La/H] et al. are close to zero. According to Table \ref{mean_abundance}, a 0.5 km s$^{-1}$ increase in the microturbulence serves to reduce the Ba abundance by about 0.3 dex without seriously affecting the La et al. abundances which are based on much weaker lines. Such an increase is formally greater than the quoted uncertainty affecting the determinatin of the microturbulence. As noted above, differences in the adopted microturbulence explain almost fully the different Ba abundances obtained for stars in Argus and Carina-Near by us and previous investigators.  

In short, the greater sensitivity of the Ba\,{\sc ii} line to microturbulence is likely a contributor to resolving the Ba puzzle. Within the classical picture of microturbulence, one might explore the effect of a height-dependent microturbulence with velocity increasing toward the top of the photosphere and the transition to the chromosphere. Of course, a more realistic modelling of microturbulence would be to adopt a 3D model atmosphere in which granulation is modelled hydrodynamically. The depth of formation of the Ba\,{\sc ii} line is not so high in the photosphere that detailed consideration of the chromosphere and the photosphere-chromsphere boundary are likely to be important for the Ba puzzle. Existing non-LTE calculations (see, for example, Korotin et al. 2015) for the Ba$^+$ ion suggest that, at least in a classical atmosphere, non-LTE line formation will not resolve the Ba puzzle (see also D'Orazi et al. 2012).
 
\section{Associations, Clusters and Field Stars}

Star formation occurring in giant molecular clouds is thought to provide two groups of stars -- open clusters and stellar associations. An OC is a collection of stars with bound by gravity. Stars in an OC have the composition and kinematics of the natal cloud and share a common age. Any gas remaining with the cluster at birth is presumed to be lost  before it can be contaminated by ejecta from the cluster's stars; indeed, supernovae from the massive stars may be responsible for cleansing the cluster of residual gas. Stars are slowly lost from a cluster to join the field star population. Small and poorly populated OCs dissolve primarily through internal dynamical effects in less than 10$^8$ Myr. Clusters of intermediate mass (say, 500 to 1000 $M_\odot$) may survive several Gyr (Pavani \& Bica 2007). 

A stellar association is a loosely bound group of young stars that fails to form a cluster at birth. Some associations may have enjoyed once a close link to an OC. In this vein, we note that De Silva et al. (2013) trace stars in the Argus association, also referred to as a moving group, to the dissolving halo of the OC IC 2391 (see also Desidera et al. 2011 and references therein). This connection is made using the high Ba abundance of association and cluster as `chemical tag'. 

In this scenario in which giant molecular clouds spawn clusters and associations and, then, stars are shed from the clusters and associations, one expects identical compositions for young OCs,  associations and young field stars in the solar neighbourhood and formed from single giant molecular cloud. This expectation is more easily stated than comprehensively tested for OCs, associations and field stars across the full range of elements from H to Eu and even beyond. 

\begin{table*}
{\fontsize{9}{9}\selectfont
\caption{Mean elemental abundance ratios, [X/Fe], for Na to Eu for our sample of five associations, the thin disc mean abundances from Reddy et al. (2012 \& 2013) and Reddy et al. (2015) for OC giants, and Luck \& Heiter (2007) for field giants, and Luck \& Heiter (2005 \& 2006) and Bensby et al. (2014) for field dwarfs, and Luck \& Lambert (2011) for Cepheids.}
\label{lit_compare}
\begin{tabular}{lcccccc}   \hline
\multicolumn{1}{l}{Species}\vline & \multicolumn{1}{c}{Associations}\vline & \multicolumn{1}{c}{LH05 \& LH06}\vline & \multicolumn{1}{c}{B14}\vline & \multicolumn{1}{c}{OC mean}\vline & \multicolumn{1}{c}{LH07} \vline & \multicolumn{1}{c}{LL11} \\ \cline{1-7} 
\multicolumn{1}{l}{ }\vline & \multicolumn{1}{c}{Dwarfs}\vline & \multicolumn{2}{c}{Field dwarfs}\vline  & \multicolumn{2}{c}{Red giants}\vline  & \multicolumn{1}{c}{Cepheids}  \\ 
% \multicolumn{1}{l}{$[$Fe/H$]$ $\sim$} & \multicolumn{1}{c}{(-0.1 to 0.1 dex)} & \multicolumn{1}{c}{(-0.1 to 0.1 dex)} & \multicolumn{1}{c}{(-0.1 to 0.1 dex)} \\ 
\hline

 $[$O/Fe$]$  & $+0.12\pm0.06$ &    $\ldots$    & $+0.01\pm0.09$ &    $\ldots$    &    $\ldots$     &  $-0.10\pm0.13$  \\
 $[$Na/Fe$]$ & $-0.01\pm0.05$ & $+0.04\pm0.10$ & $+0.00\pm0.05$ & $+0.26\pm0.07$ & $+0.11\pm0.06$  &  $+0.17\pm0.08$  \\ 
 $[$Mg/Fe$]$ & $-0.04\pm0.03$ & $+0.14\pm0.13$ & $+0.01\pm0.10$ & $+0.04\pm0.05$ & $+0.07\pm0.12$  &  $-0.08\pm0.16$  \\
 $[$Al/Fe$]$ & $-0.05\pm0.05$ & $+0.04\pm0.10$ & $+0.03\pm0.08$ & $ 0.00\pm0.07$ & $+0.08\pm0.05$  &  $+0.09\pm0.07$  \\
 $[$Si/Fe$]$ & $+0.04\pm0.02$ & $+0.03\pm0.06$ & $+0.02\pm0.04$ & $+0.16\pm0.06$ & $+0.12\pm0.05$  &  $+0.05\pm0.05$  \\
 $[$Ca/Fe$]$ & $+0.04\pm0.01$ & $+0.02\pm0.07$ & $+0.03\pm0.04$ & $+0.07\pm0.06$ & $-0.05\pm0.05$  &  $-0.05\pm0.08$  \\
 $[$Ti/Fe$]$ & $+0.01\pm0.01$ & $+0.05\pm0.09$ & $+0.02\pm0.04$ & $+0.01\pm0.05$ & $-0.01\pm0.03$  &  $+0.03\pm0.08$  \\
 $[$Cr/Fe$]$ & $+0.00\pm0.05$ & $+0.03\pm0.06$ & $+0.00\pm0.03$ & $+0.04\pm0.03$ & $+0.02\pm0.04$  &  $ 0.00\pm0.07$  \\
 $[$Mn/Fe$]$ & $+0.02\pm0.04$ & $-0.03\pm0.10$ &     $\ldots$   & $-0.08\pm0.06$ & $+0.11\pm0.09$  & $-0.05\pm0.09$  \\
%  $[$Co/Fe$]$ & $-0.08\pm0.03$ &   $\ldots$      &     $\ldots$     \\
 $[$Ni/Fe$]$ & $-0.05\pm0.01$ & $0.00\pm0.04$  & $-0.02\pm0.03$ & $-0.01\pm0.04$ & $+0.01\pm0.04$  &  $-0.05\pm0.04$  \\
 $[$Zn/Fe$]$ & $-0.06\pm0.06$ & $-0.03\pm0.11$ & $ 0.00\pm0.11$ & $-0.14\pm0.14$ &   $\ldots$      &  $+0.05\pm0.24$  \\
 $[$Y/Fe$]$  & $+0.06\pm0.02$ & $-0.06\pm0.18$ & $-0.01\pm0.13$ & $+0.10\pm0.05$ & $+0.06\pm0.11$  &  $+0.14\pm0.07$  \\
 $[$Zr/Fe$]$ & $ 0.00\pm0.00$ & $-0.33\pm0.34$ & $-0.01\pm0.13$ & $+0.13\pm0.11$ &   $\ldots$      &  $+0.01\pm0.12$  \\
 $[$Ba/Fe$]$ & $+0.21\pm0.09$ & $-0.02\pm0.11$ & $+0.05\pm0.16$ & $+0.23\pm0.21$ & $-0.02\pm0.18$  &  $\bf+0.17\pm0.32^{\bf A14}$  \\
 $[$La/Fe$]$ & $+0.04\pm0.04$ & $+0.08\pm0.11$ &    $\ldots$    & $+0.16\pm0.06$ &    $\ldots$     &  $+0.18\pm0.10$  \\
 $[$Ce/Fe$]$ & $0.00\pm0.04$  & $+0.06\pm0.12$ &    $\ldots$    & $+0.22\pm0.10$ & $+0.03\pm0.11$  &  $-0.02\pm0.12$  \\
 $[$Nd/Fe$]$ & $+0.02\pm0.06$ & $+0.03\pm0.14$ &    $\ldots$    & $+0.18\pm0.07$ & $-0.03\pm0.07$  &  $+0.07\pm0.11$  \\
 $[$Sm/Fe$]$ & $+0.05\pm0.05$ & $-0.10\pm0.14$ &    $\ldots$    & $+0.16\pm0.07$ &    $\ldots$     &  $-0.03\pm0.14$  \\
 $[$Eu/Fe$]$ & $ 0.00\pm0.01$ & $+0.15\pm0.08$ &    $\ldots$    & $+0.14\pm0.07$ & $+0.06\pm0.08$  &  $+0.08\pm0.09 $  \\

\hline
\end{tabular}  }
\flushleft
Note: LH07= Luck \& Heiter (2007); LH05 \& LH06= Luck \& Heiter (2005 \& 2006); B14= Bensby et al. (2014), LL11= Luck \& Lambert (2011) and A14= Andrievsky, Luck \& Korotin (2014). 
\end{table*}

The primary stimulus for this paper was the reported increase in the Ba abundance in OCs with decreasing age such that in the youngest clusters Ba was claimed to reach [Ba/Fe] $\simeq +0.6$ (D'Orazi et al. 2009). The Ba enrichment in associations is lower than the high [Ba/Fe] reported for the youngest clusters; our analysis confirms and extends D'Orazi et al.'s (2012) analysis of heavy elements in associations to show that [Ba/Fe] has a mean value of about $+0.2$ with [Fe/H] $\simeq 0.0$ with other heavy elements averaging about zero. These analyses of cluster and association members use stars of the same spectral types (mainly F and G dwarfs) and similar, almost identical, standard methods of analysis such that systematic differences between authors are surely small. Thus, taken at face value, the expectation of identical compositions for local OCs and associations
is not confirmed for Ba. (Note, however, that De Silva et al. (2013) claimed [Ba/Fe] $= +0.6$ for the Argus association, a value not confirmed here.) We have suggested in the previous section that the demonstration that Ba differs from other examined heavy elements in associations may be due to its representation by strong lines in spectra of F-G dwarfs coupled with failures of current stellar atmospheres and line formation theory which have a smaller, even negligible, effect on the weak lines of other heavy elements. This conclusion should now be tested by abundance analyses of F-G dwarfs in the youngest local OCs including elements from Y to Eu and, of course, Ba.

There exists an extensive literature on heavy element abundances in OCs drawn from dwarfs and/or red giants but little of the available data concerns the very youngest clusters. Furthermore, few of the clusters are local and many were chosen because they had a non-solar metallicity and, thus, extracting the appropriate abundances for comparison with the abundances of local associations with [Fe/H] $\simeq 0$ is a non-trivial exercise over
and above establishing and correcting for systematic differences between abundances obtained from F-G dwarfs, as in the local associations, and G-K red giants, as in many of the OCs. Recent studies reporting heavy element abundances include Maiorca et al. (2011), Reddy et al. (2012, 2013, 2015), Jacobson \& Friel (2013) and Mishenina et al. (2015) with several of these papers including compilations drawn from from the literature. 

Mishenina et al. assemble Y, Ba and La abundances from OCs. Many of the OCs have ages less than 1 Gyr but very few have ages less than 0.5 Gyr. By extrapolation of their plots of [X/Fe] versus age (their Figure 4) to zero age, one obtains [Y/Fe] $\simeq+0.1\pm0.1$, [Ba/Fe] $\simeq 0.4\pm0.3$, and [La/Fe] $\simeq +0.1\pm0.2$. These extrapolations are consistent with values obtained from Jacobson \& Friel's smaller sample of OCs: [Zr/Fe] $= +0.2$, [Ba/Fe] $= +0.4$ and [La/Fe] $= 0.0$ where one may anticipate [Y/Fe] should be very similar to [Zr/Fe]. Jacobson \& Friel also obtained [Eu/Fe] $=-0.15$ at zero age. The {\it very} much larger spread in [Ba/Fe] than for [Y/Fe] and [La/Fe] and extension of the Ba spread to older OCs  suggests that the Ba abundances from the strong Ba\,{\sc ii} lines  in spectra of red giants are subject to large systematic errors from, as yet unknown sources. This suggestion is supported by measurements of Ba abundances for field F-G dwarfs in the Galactic thin disc (Luck \& Heiter 2005, 2006; Reddy et al. 2003, 2006; Mishenina et al. 2013; Bensby et al. 2014) which show no measureable scatter in [Ba/Fe] at a given [Fe/H] or a given age and scatter about [Ba/Fe] $\simeq 0.0$. Oddly, a survey of Ba in local field G-K giants (Luck \& Heiter 2007) gives a mean [Ba/Fe] $=-0.02\pm0.18$ for [Fe/H] $\simeq 0.0$ stars with no indication of the large scatter suggested by Ba results for OCs. A real spread in heavy element abundances, i.e., [X/Fe], at [Fe/H] $\simeq 0$ for OCs is suggested by Reddy et al. (2015) whose determinations of [X/Fe] for a given cluster had uncertainties of about $\pm0.05$ but the values from cluster to cluster were spread over a range of up to $+$0.4 dex from [X/Fe] $\simeq 0.0$ and were correlated approximately with the $s$-process contribution, i.e., the spread for [Ce/Fe] was larger than that for [Eu/Fe]. Results for the local associations fall at the low end of the spreads.  

Another point of comparison for stars in associations and young clusters are young field stars. In this respect, Cepheids are available with their ages comparable to those of associations. Abundances for many elements in a large sample of Cepheids were provided by Luck \& Lambert (2011) with Ba 
obtained from the same spectra by Andrievsky, Luck \& Korotin 2014). Mean results for the Cepheids with [Fe/H] across the interval [Fe/H]  from $-0.10$ to $+0.10$ are given in Table \ref{lit_compare}. The close agreement between [X/Fe] from the associations with values from Cepheids is both pleasing and surprising. Surprise is occasioned by the fact that both analyses are rooted in the LTE approximation (and other assumptions) but Cepheids as warm supergiants differ greatly from F-G dwarfs so that noticeable systematic effects would seem to be inevitable.  

Table \ref{lit_compare} summarizes the compositions of the local associations and the comparison samples mentioned above. Entries for the field dwarfs are taken from Luck \& Heiter (2005, 2006) and from Benbsy et al. (2014) with the constraint that [Fe/H] is in the range $-0.10$ to $+0.10$. Almost all of the selected stars are from the Galactic thin disc. For local field red giants, again with the constraint that [Fe/H] fall within the interval $-0.10$ to $+0.10$, stars were chosen from Luck \& Heiter (2007). For OCs, red giants were taken from the following clusters: NGC 1817 (Reddy et al. 2012), NGC 2251, 2482, 2527, 2539 (Reddy et al. 2013), NGC 1342, 1662, 1912, 2447 (Reddy et al. 2015) and NGC 3114 (Santrich et al. 2013). The OC sample covers the age range 160-450 Myr and [Fe/H] from $-0.14$ to 0.0 and heliocentric distances of 400-2000 pc. This OC sample appears in Table \ref{lit_compare} to be the exception with respect to heavy element abundances but, as noted above, the heavy element abundances in the OC sample span a range and the results for the local associations fall at one end of the range in [X/Fe] shown by the OCs. All other samples have element abundances which compare favourably with abundances for the local associations with the sole exception of Ba where [Ba/Fe] $\simeq +0.2$ for the F-G dwarfs in the associations but [Ba/Fe] $\simeq 0.0$ for field dwarfs, giants and Cepheids. Table \ref{lit_compare}, thus, reinforces the impression that Ba in the young F-G dwarfs of local associations may have an overestimated abundance and because its abundance is not overestimated in field dwarfs and giants one may suspect that an explanation for the Ba overestimation is to be found in the interaction in young stars between stellar activity and the stellar photosphere. Such an interaction is lessened in older stars as the level of stellar activity decays. 

\section{Concluding Remarks}

In this paper, we have presented elemental abundances from Li to Eu for F-G dwarfs belonging to five nearby stellar associations (Argus, Carina-Near, Hercules-Lyra, Orion and Subgroup B4). Stars in a given association have very similar compositions and all five associations with [Fe/H] ranging from slightly subsolar to solar share very similar [X/Fe] with Ba as the sole striking exception. The [X/Fe] $\simeq 0.0$ for all measured heavy elements (Y, Zr, La, Ce, Nd, Sm and Eu) except for Ba ([Ba/Fe] attains $+0.3$ in the Argus association). This isolation of Ba among the set of heavy elements prompts the suggestion that the Ba abundance provided necessarily by strong Ba\,{\sc ii} lines is systematically overestimated by standard methods of abundance analysis. 

The study reported here was prompted by the report that [Ba/Fe] in young clusters increased with decreasing age attaining [Ba/Fe] $\simeq +0.6$ in clusters with ages of about 50 Myr (D'Orazi et al. 2009). Associations of comparable age show lower [Ba/Fe] (D'Orazi et al. 2012; this paper). This
discrepancy between the youngest clusters and associations warrants a close reinvestigation by high-resolution spectroscopy. 

\vskip1ex 
{\bf Acknowledgements:}
 
We are grateful to the McDonald Observatory's Time Allocation Committee for granting us observing time for this project. DLL wishes to thank the Robert A. Welch Foundation of Houston, Texas for support through grant F-634. We thank Maurizio Busso for helpful insights into operation of the $s$-process. We are grateful to the anonymous referee for a very careful and constructive report that led to improvements of the Paper.

This research has made use of the WEBDA database, operated at the Institute for Astronomy of the University of Vienna and the NASA ADS, USA. 
This research has made use of the SIMBAD database and Aladin sky atlas, operated at CDS, Strasbourg, France. This publication makes use of data products from the Two Micron All Sky Survey, which is a joint project of the University of Massachusetts and the Infrared Processing and Analysis Center/California Institute of Technology, funded by the National Aeronautics and Space Administration (NASA) and the National Science Foundation (NSF).

\vspace{1cm} 
\begin{table*} 
 {\fontsize{8.5}{8.5}\selectfont
\caption[Elemental abundances for stars in the Argus \& Carina-Near]{Elemental abundance ratios, [X/Fe], for individual stars in Argus and Carina-Near associations. The abundances calculated by synthesis are presented in bold typeface while the remaining elemental abundances were calculated using the line EWs. Numbers in the parentheses indicate the number of lines used in calculating the abundance of that element. The average abundance is not tabulated for the Ba {\scs I} line as it is affected severely by the departures from LTE.} 
% \vspace{0.2cm}
\label{abu_Argus}
\begin{tabular}{lcccccc}   \hline
\multicolumn{1}{l}{Species} & \multicolumn{1}{c}{HD 61005} & \multicolumn{1}{c}{BD-20 2977} &\multicolumn{1}{c}{Argus$_{\mbox{Avg.}}$} \vline & \multicolumn{1}{c}{HD 103742} & \multicolumn{1}{c}{HD 103743} & \multicolumn{1}{c}{Carina-Near$_{\mbox{Avg.}}$}  \\ \hline

log\,$\epsilon$(Li) &  $\bf+3.10$  &  $\bf+3.55$      &  $\ldots$       &   $\bf+2.95$      & $\bf+2.95$        &  $\ldots$     \\
$[$O I/Fe$]$  &  $\bf+0.06$      &$+0.13\pm0.02$(2) &$+0.09\pm0.02$   & $+0.14\pm0.03$(4) &$+0.08\pm0.02$(3)  &$+0.11\pm0.02$   \\
$[$Na I/Fe$]$ &$+0.03\pm0.04$(4) &$+0.11\pm0.04$(5) &$+0.07\pm0.03$   & $-0.06\pm0.03$(5) &$-0.04\pm0.02$(3)  &$-0.05\pm0.02$   \\
$[$Mg I/Fe$]$ &$-0.10\pm0.02$(4) &$-0.07\pm0.03$(4) &$-0.08\pm0.02$   & $-0.06\pm0.03$(3) &$-0.05\pm0.00$(1)  &$-0.05\pm0.02$   \\
$[$Al I/Fe$]$ &$-0.03\pm0.04$(5) &$-0.02\pm0.02$(5) &$-0.02\pm0.02$   & $-0.07\pm0.03$(5) &$-0.12\pm0.05$(5)  &$-0.09\pm0.03$   \\
$[$Si I/Fe$]$ &$+0.04\pm0.02$(9) &$-0.05\pm0.04$(11)&$ 0.00\pm0.02$   & $+0.03\pm0.05$(14)&$+0.04\pm0.04$(10) &$+0.03\pm0.03$   \\
$[$Ca I/Fe$]$ &$+0.06\pm0.04$(10)&$+0.07\pm0.05$(9) &$+0.06\pm0.03$   & $+0.04\pm0.04$(12)&$+0.05\pm0.04$(10) &$+0.04\pm0.03$   \\
$[$Sc I/Fe$]$ &$-0.08\pm0.02$(3) &$+0.05\pm0.00$(1) &$-0.01\pm0.01$   & $ 0.00\pm0.04$(5) &$-0.04\pm0.05$(2)  &$-0.02\pm0.03$   \\
% $[$Sc II/Fe$]$&$+0.03\pm0.04$(4) &$-0.05\pm0.01$(2) &$-0.01\pm0.02$   & $-0.01\pm0.03$(3) &$-0.02\pm0.04$(3)  &$-0.01\pm0.03$   \\ 
$[$Sc II/Fe$]$&  $\bf+0.03$      &   $\bf+0.08$     &  $\bf+0.05$     &  $\bf+0.01$       &   $\bf+0.11$      &  $\bf+0.06$     \\
$[$Ti I/Fe$]$ &$ 0.00\pm0.03$(17) &$+0.01\pm0.04$(11)&$ 0.00\pm0.02$  & $ 0.00\pm0.04$(14)&$+0.04\pm0.03$(9)  &$+0.02\pm0.03$   \\
$[$Ti II/Fe$]$&$+0.03\pm0.03$(4) &$+0.01\pm0.04$(3) &$+0.02\pm0.02$   & $-0.04\pm0.05$(6) &$-0.06\pm0.04$(5)  &$-0.05\pm0.03$   \\
$[$V I/Fe$]$  &$+0.03\pm0.05$(10)&$+0.06\pm0.03$(9) &$+0.04\pm0.03$   & $ 0.00\pm0.03$(7) &$+0.03\pm0.02$(4)  &$+0.01\pm0.02$   \\
$[$Cr I/Fe$]$ &$+0.03\pm0.04$(12) &$+0.08\pm0.04$(10)&$+0.05\pm0.03$  & $+0.02\pm0.04$(14)&$+0.05\pm0.04$(9)  &$+0.03\pm0.03$   \\ 
$[$Cr II/Fe$]$&$+0.04\pm0.03$(4)  &$ 0.00\pm0.04$(4) &$+0.02\pm0.02$  & $+0.01\pm0.03$(8) &$+0.02\pm0.02$(6)  &$+0.01\pm0.02$   \\ 
$[$Mn I/Fe$]$ & $\bf+0.05$        &   $\bf+0.09$     &  $\bf+0.06$    &  $\bf-0.03$       &   $\bf+0.08$       &  $\bf+0.02$    \\
$[$Fe I/H$]$  &$-0.01\pm0.04$(139)&$-0.01\pm0.04$(92)&$-0.01\pm0.03$  & $-0.02\pm0.03$(119)&$+0.03\pm0.03$(128)&$ 0.00\pm0.02$  \\
$[$Fe II/H$]$ &$+0.01\pm0.04$(13)&$ 0.00\pm0.03$(7)  &$ 0.00\pm0.02$  & $-0.02\pm0.03$(14) &$+0.02\pm0.04$(14) &$ 0.00\pm0.03$  \\ 
$[$Co I/Fe$]$ &$-0.20\pm0.03$(6) &$-0.05\pm0.00$(1)  &$-0.12\pm0.01$  & $-0.07\pm0.03$(6) &$-0.14\pm0.03$(5)  &$-0.10\pm0.02$   \\ 
$[$Ni I/Fe$]$ &$-0.08\pm0.04$(19)&$-0.07\pm0.04$(15) &$-0.07\pm0.03$  & $-0.05\pm0.04$(30)&$-0.04\pm0.04$(21) &$-0.04\pm0.03$   \\ 
% $[$Cu I/Fe$]$ &$-0.05\pm0.02$(2) &$-0.07\pm0.01$(2)  &$-0.06\pm0.01$  &$-0.13\pm0.02$(2)  &$-0.11\pm0.04$(2)  &$-0.12\pm0.02$   \\
% $[$Cu I/Fe$]$ &   $\bf-0.25$     &  $\bf-0.30$       &  $\bf-0.27$    &    $\bf-0.28$     &  $\bf-0.32$       &  $\bf-0.30$     \\
$[$Zn I/Fe$]$ &   $\bf-0.08$     &  $\bf-0.07$       &  $\bf-0.07$    &    $\bf-0.10$     &  $\bf-0.10$       &  $\bf-0.10$     \\ 
$[$Y II/Fe$]$ &$+0.07\pm0.04$(4) &$+0.07\pm0.04$(2) &$+0.07\pm0.03$   & $+0.04\pm0.05$(4) &$+0.09\pm0.02$(3)  &$+0.06\pm0.03$   \\
$[$Zr I/Fe$]$ &$-0.04\pm0.01$(2) &$ 0.00\pm0.00$(1) &$-0.02\pm0.00$   & $+0.05\pm0.03$(3) &$+0.01\pm0.03$(3)  &$+0.03\pm0.02$   \\
$[$Ba I/Fe$]$ &  $\bf-0.25 $     & $\bf-0.25$       &   $\ldots$      &   $\bf+0.02$      &  $\bf-0.02$       &   $\ldots$    \\
$[$Ba II/Fe$]$&  $\bf+0.30 $     & $\bf+0.35$       &   $\bf+0.32$    &   $\bf+0.32$      &  $\bf+0.32$       &   $\bf+0.32$    \\
$[$La II/Fe$]$&$+0.01\pm0.00$(1) &$+0.07\pm0.00$(1) &$+0.04\pm0.00$   & $+0.01\pm0.05$(2) &  $-0.01\pm0.03$(2)&$ 0.00\pm0.03$   \\
$[$Ce II/Fe$]$&$-0.01\pm0.00$(1) &   $ $            &$-0.01\pm0.00$   & $-0.01\pm0.00$(2) &$-0.02\pm0.00$(1)  &$-0.01\pm0.00$   \\
$[$Nd II/Fe$]$&   $\ldots$       &  $\ldots$        &   $\ldots$      & $+0.10\pm0.03$(4) &   $  $            &$+0.10\pm0.03$   \\
$[$Sm II/Fe$]$&   $\bf+0.05$     &  $\bf 0.00$      &  $\bf+0.02$     & $+0.03\pm0.03$(3) &   $\bf-0.02$      &$ 0.00\pm0.03$   \\
$[$Eu II/Fe$]$&   $\bf0.00$      &  $\bf0.00$       &  $\bf 0.00$     &  $\bf+0.02$       &   $\bf-0.02$      &  $\bf0.00$      \\

\hline
\end{tabular} } 
\end{table*}
% \end{landscape}

\begin{table*} 
{\fontsize{8.5}{8.5}\selectfont
\caption[Elemental abundances for stars in the Orion \& Subgroup B4]{Same as Table \ref{abu_Argus} but for individual stars in associations Orion and Subgroup B4.} 
\label{abu_Orion}
\begin{tabular}{lcccccc}   \hline
\multicolumn{1}{l}{Species} & \multicolumn{1}{c}{Parenago 2339} & \multicolumn{1}{c}{Parenago 2374} &\multicolumn{1}{c}{Orion$_{\mbox{Avg.}}$} \vline & \multicolumn{1}{c}{HD 113449} & \multicolumn{1}{c}{HD 152555} & \multicolumn{1}{c}{Subgroup B4$_{\mbox{Avg.}}$}  \\ \hline

log\,$\epsilon$(Li) &  $\bf+3.00$  & $\bf+2.40$        &  $\ldots$      &  $\bf+2.55$        & $\bf+3.25$        &  $\ldots$      \\
$[$O I/Fe$]$  &$+0.19\pm0.03$(3) &$+0.25\pm0.04$(4)  &$+0.22\pm0.03$  & $+0.15\pm0.03$(2)  &$+0.03\pm0.01$(3)  &$+0.09\pm0.01$   \\
$[$Na I/Fe$]$ &$+0.02\pm0.03$(3) &$ 0.00\pm0.05$(6)  &$+0.01\pm0.03$  & $-0.07\pm0.02$(5)  &$-0.05\pm0.01$(2)  &$-0.06\pm0.01$   \\
$[$Mg I/Fe$]$ &$-0.02\pm0.02$(2) &$+0.04\pm0.00$(2)  &$+0.01\pm0.01$  & $-0.05\pm0.03$(4)  &$-0.02\pm0.00$(1)  &$-0.03\pm0.02$   \\
$[$Al I/Fe$]$ &$ 0.00\pm0.03$(4) &$+0.05\pm0.02$(4)  &$+0.02\pm0.02$  & $-0.14\pm0.03$(7)  &$-0.03\pm0.02$(3)  &$-0.08\pm0.02$   \\
$[$Si I/Fe$]$ &$+0.11\pm0.03$(6) &$+0.03\pm0.05$(12) &$+0.07\pm0.03$  & $+0.01\pm0.04$(14) &$+0.09\pm0.02$(5)  &$+0.05\pm0.02$   \\
$[$Ca I/Fe$]$ &$+0.04\pm0.03$(5) &$+0.03\pm0.05$(11) &$+0.03\pm0.03$  & $+0.01\pm0.04$(12) &$+0.05\pm0.04$(6)  &$+0.03\pm0.03$   \\
% $[$Sc II/Fe$]$&  $\ldots$        &$+0.10\pm0.06$(3)  &$+0.10\pm0.06$  & $-0.06\pm0.04$(3)  &   $\ldots$        &$-0.06\pm0.04$   \\ 
$[$Sc II/Fe$]$&  $\bf-0.09$      &   $\bf+0.12$      &  $\bf+0.01$    &   $\bf+0.03$       &   $\bf+0.04$      &  $\bf+0.03$     \\
$[$Ti I/Fe$]$ &$-0.02\pm0.04$(3) &$+0.05\pm0.04$(8)  &$+0.01\pm0.03$  & $-0.10\pm0.03$(12) &$+0.09\pm0.02$(4)  &$ 0.00\pm0.02$   \\
$[$Ti II/Fe$]$&$-0.03\pm0.00$(1) &$+0.01\pm0.03$(4)  &$-0.01\pm0.02$  & $-0.10\pm0.02$(5)  &$-0.03\pm0.04$(3)  &$-0.06\pm0.02$   \\
$[$V I/Fe$]$  &$+0.08\pm0.00$(1) &$ 0.00\pm0.05$(3)  &$+0.04\pm0.03$  & $-0.08\pm0.02$(7)  &$+0.15\pm0.04$(2)  &$+0.03\pm0.02$   \\
$[$Cr I/Fe$]$ &$-0.13\pm0.04$(5) &$-0.02\pm0.03$(10) &$-0.07\pm0.03$  & $-0.03\pm0.03$(13) &$-0.08\pm0.05$(4)  &$-0.05\pm0.03$   \\ 
$[$Cr II/Fe$]$&$-0.10\pm0.01$(2) &$+0.07\pm0.03$(5)  &$-0.01\pm0.02$  & $+0.05\pm0.03$(7)  &$-0.08\pm0.05$(4)  &$-0.01\pm0.03$   \\ 
$[$Mn I/Fe$]$ & $\bf-0.01$       &   $\bf-0.04$      & $\bf-0.02$     & $\bf-0.03$         &  $\bf-0.06$       &  $\bf-0.04$     \\
$[$Fe I/H$]$  &$-0.14\pm0.04$(19)&$-0.07\pm0.04$(115)&$-0.10\pm0.03$  & $-0.09\pm0.04$(136)&$-0.04\pm0.03$(63) &$-0.06\pm0.03$   \\
$[$Fe II/H$]$ &$-0.14\pm0.04$(7) &$-0.06\pm0.04$(14) &$-0.10\pm0.03$  & $-0.07\pm0.03$(12) &$-0.03\pm0.05$(4)  &$-0.05\pm0.03$   \\ 
$[$Co I/Fe$]$ &$-0.08\pm0.00$(1) &$-0.02\pm0.04$(2)  &$-0.05\pm0.02$  & $-0.07\pm0.02$(6)  &$-0.08\pm0.03$(3)  &$-0.07\pm0.02$   \\ 
$[$Ni I/Fe$]$ &$-0.13\pm0.03$(7) &$+0.01\pm0.04$(20) &$-0.06\pm0.03$  & $-0.08\pm0.04$(22) &$-0.03\pm0.04$(7)  &$-0.05\pm0.03$   \\ 
% $[$Cu I/Fe$]$ &   $\ldots$       &$-0.01\pm0.05$(2)  &$-0.01\pm0.05$  &$-0.01\pm0.04$(3)   &$-0.08\pm0.00$(1)  &$-0.04\pm0.02$   \\
% $[$Cu I/Fe$]$ &   $\ldots$       & $\bf-0.09$        &  $\bf-0.09$    &  $\bf-0.08$        &  $\bf-0.26$       &  $\bf-0.17$     \\
$[$Zn I/Fe$]$ &   $\ldots$       & $\bf+0.06$        &  $\bf+0.06$    &  $\bf-0.03$        &  $\bf-0.18$       &  $\bf-0.11$     \\ 
$[$Y II/Fe$]$ &$+0.04\pm0.01$(2) &$-0.02\pm0.05$(2)  &$+0.02\pm0.03$  & $+0.07\pm0.01$(4)  &$+0.07\pm0.04$(2)  &$+0.07\pm0.02$   \\
$[$Zr I/Fe$]$ &  $\ldots$        &  $\ldots$         &  $\ldots$      & $-0.04\pm0.05$(5)  &    $\ldots$       &$-0.04\pm0.05$   \\
$[$Ba I/Fe$]$ &  $\bf+0.10$      & $\bf+0.05$        &  $\ldots$    &   $\bf-0.51$       &  $\bf+0.11$       &   $\ldots$    \\
$[$Ba II/Fe$]$&  $\bf+0.09$      & $\bf+0.06$        &  $\bf+0.07$    &   $\bf+0.22$       &  $\bf+0.24$       &   $\bf+0.23$    \\
$[$La II/Fe$]$&  $\ldots$        &$+0.06\pm0.04$(2)  &$+0.06\pm0.04$  & $-0.01\pm0.01$(3)  &    $\ldots$       &$-0.01\pm0.01$   \\
$[$Ce II/Fe$]$&  $\ldots$        &$-0.05\pm0.03$(2)  &$-0.05\pm0.03$  & $+0.05\pm0.01$(2)  &    $\ldots$       &$+0.05\pm0.01$   \\
$[$Nd II/Fe$]$&  $\ldots$        &$-0.06\pm0.03$(3)  &$-0.06\pm0.03$  & $+0.08\pm0.03$(3)  &    $\ldots$       &$+0.08\pm0.03$   \\
$[$Sm II/Fe$]$&  $\ldots$        &$+0.11\pm0.02$(3)  &$+0.11\pm0.02$  & $+0.11\pm0.03$(6)  &    $\ldots$       &$+0.11\pm0.03$   \\
$[$Eu II/Fe$]$&  $\ldots$        &  $\bf+0.01$       &  $\bf+0.01$    &    $\ldots$        &    $\ldots$       &   $\ldots$      \\

\hline
\end{tabular}  } 
\end{table*}
% \end{landscape}

\begin{table*}
{\fontsize{8.5}{8.5}\selectfont
\caption[Elemental abundances for stars in the association Hercules-Lyra]{Same as Table \ref{abu_Argus} but for individual stars in the Hercules-Lyra association.} 
% \vspace{0.2cm}
\label{abu_Hercules}
\begin{tabular}{lccccc}   \hline
\multicolumn{1}{c}{Species}& \multicolumn{1}{c}{HD 166}& \multicolumn{1}{c}{HD 10008}& \multicolumn{1}{c}{HD 17925} & \multicolumn{1}{c}{Hercules-Lyra$_{\mbox{Avg.}}$} \\ \hline

log\,$\epsilon$(Li) &   $\bf+2.30$  &   $\bf+2.35$      &    $\bf+2.35$    &  $\ldots$      \\
$[$O I/Fe$]$  &$+0.02\pm0.04$(4)  &$+0.06\pm0.03$(4)  &$+0.06\pm0.04$(4) &$+0.05\pm0.02$    \\
$[$Na I/Fe$]$ &$ 0.00\pm0.02$(5)  &$-0.10\pm0.04$(4)  &$ 0.00\pm0.02$(5) &$-0.03\pm0.02$    \\
$[$Mg I/Fe$]$ &$-0.04\pm0.02$(4)  &$-0.10\pm0.03$(4)  &$-0.08\pm0.04$(4) &$-0.07\pm0.02$    \\
$[$Al I/Fe$]$ &$-0.09\pm0.02$(5)  &$-0.16\pm0.04$(5)  &$-0.05\pm0.05$(5) &$-0.10\pm0.02$    \\
$[$Si I/Fe$]$ &$+0.03\pm0.04$(13) &$ 0.00\pm0.04$(13) &$+0.07\pm0.04$(13)&$+0.03\pm0.02$    \\
$[$Ca I/Fe$]$ &$+0.05\pm0.04$(11) &$-0.01\pm0.05$(12) &$+0.03\pm0.05$(11) &$+0.02\pm0.03$   \\
$[$Sc I/Fe$]$ &$-0.08\pm0.03$(2)  &$-0.05\pm0.05$(2)  &$ 0.00\pm0.06$(2)  &$-0.04\pm0.03$   \\
% $[$Sc II/Fe$]$&$+0.01\pm0.05$(3)  &$-0.02\pm0.02$(3)  &$-0.01\pm0.09$(3)  &$-0.01\pm0.03$   \\ 
$[$Sc II/Fe$]$& $\bf+0.05$        & $\bf-0.01$        &  $\bf-0.02$       & $\bf+0.01$      \\
$[$Ti I/Fe$]$ &$+0.04\pm0.03$(13) &$-0.01\pm0.03$(12) &$-0.01\pm0.04$(11) &$+0.01\pm0.02$   \\
$[$Ti II/Fe$]$&$-0.03\pm0.02$(5)  &$-0.03\pm0.04$(6)  &$-0.05\pm0.04$(5)  &$-0.04\pm0.02$   \\
$[$V I/Fe$]$  &$+0.05\pm0.03$(7)  &$ 0.00\pm0.05$(7)  &$+0.06\pm0.05$(6)  &$+0.04\pm0.03$   \\
$[$Cr I/Fe$]$ &$+0.07\pm0.03$(12) &$-0.01\pm0.04$(13) &$+0.02\pm0.04$(12) &$+0.03\pm0.02$   \\ 
$[$Cr II/Fe$]$&$+0.04\pm0.03$(8)  &$+0.06\pm0.03$(8)  &$+0.12\pm0.04$(5)  &$+0.07\pm0.02$   \\ 
$[$Mn I/Fe$]$ &   $\bf+0.05$      &  $\bf-0.01$       &  $\bf+0.13$       &  $\bf+0.06$     \\
$[$Fe I/H$]$  &$+0.06\pm0.04$(144)&$-0.02\pm0.04$(143)&$+0.02\pm0.04$(133)&$+0.02\pm0.02$   \\
$[$Fe II/H$]$ &$+0.05\pm0.03$(15) &$-0.01\pm0.04$(14) &$+0.04\pm0.04$(14) &$+0.03\pm0.02$   \\ 
$[$Co I/Fe$]$ &$-0.04\pm0.04$(5)  &$-0.08\pm0.02$(5)  &$-0.04\pm0.02$(6)  &$-0.05\pm0.02$   \\ 
$[$Ni I/Fe$]$ &$-0.02\pm0.04$(27) &$-0.07\pm0.04$(26) &$-0.04\pm0.04$(25) &$-0.04\pm0.02$   \\
% $[$Cu I/Fe$]$ &$-0.05\pm0.04$(3)  &$-0.05\pm0.05$(3)   &$-0.07\pm0.01$(2) &$-0.06\pm0.02$   \\
% $[$Cu I/Fe$]$ &  $\bf-0.15$       &  $\bf-0.09$       &  $\bf-0.18$       &$\bf-0.14$       \\
$[$Zn I/Fe$]$ &  $\bf-0.08$       &  $\bf-0.09$       &  $\bf-0.11$       &$\bf-0.09$       \\ 
$[$Y II/Fe$]$ &$-0.03\pm0.04$(6)  &$+0.08\pm0.03$(6)  &$+0.17\pm0.05$(3)  &$+0.07\pm0.02$   \\
$[$Zr I/Fe$]$ &$ 0.00\pm0.03$(4)  &$+0.02\pm0.03$(6)  &$+0.07\pm0.04$(6)  &$+0.03\pm0.02$   \\
$[$Zr II/Fe$]$&$-0.01\pm0.05$(2)  &   $\ldots$        &   $\ldots$        &$-0.01\pm0.05$   \\
$[$Ba I/Fe$]$ &   $\bf-0.21$      &  $\bf-0.32$       &   $\bf-0.52$      &  $\ldots$     \\
$[$Ba II/Fe$]$&   $\bf+0.10$      &  $\bf+0.21$       &   $\bf+0.13$      &  $\bf+0.15$     \\
$[$La II/Fe$]$&$+0.06\pm0.01$(3)  &$+0.09\pm0.02$(3)  &$+0.11\pm0.04$(4)  &$+0.09\pm0.01$   \\
$[$Ce II/Fe$]$&$-0.01\pm0.00$(2)  &$+0.02\pm0.03$(3)  &$-0.01\pm0.05$(2)  &$ 0.00\pm0.02$   \\
$[$Nd II/Fe$]$&$-0.01\pm0.05$(3)  &$+0.06\pm0.04$(3)  &$-0.01\pm0.04$(2)  &$+0.01\pm0.03$   \\
$[$Sm II/Fe$]$&$-0.02\pm0.04$(5)  &$+0.02\pm0.05$(6)  &$+0.03\pm0.03$(6)  &$+0.01\pm0.02$   \\
$[$Eu II/Fe$]$&  $\bf-0.05$       &   $\bf+0.01$      &   $\bf+0.02$      &  $\bf-0.01$     \\

\hline
\end{tabular} }
\end{table*}

\end{document}